\documentclass{aa}
\usepackage{graphics}
\usepackage{natbib}
\usepackage{txfonts}
\input{psfig}
\bibpunct[; ]{(}{)}{,}{a}{}{,}
\begin{document}

%
\title{Star-formation in NGC 4038/4039 from broad- and narrow band
photometry: Cluster Destruction?\thanks{Based on observations collected
at the European Southern Observatory, Chile, programme identification
numbers 63.N-0528, 65.N-0577, and 67.B-0504}}

\titlerunning{Star Formation in NGC 4038/4039}

\author{Sabine Mengel$^1$, Matthew D. Lehnert$^2$, Niranjan Thatte$^3$,
and Reinhard Genzel$^{2,4}$}

\authorrunning{Mengel et al.}

\offprints{S. Mengel}

\institute{European Southern Observatory, Karl-Schwarzschild-Str. 2, D-85748 Garching, Germany\\
        \and
           Max-Planck-Institut f\"ur extraterrestrische Physik,
           Giessenbachstra\ss e, D-85748 Garching, Germany\\
        \and
	   University of Oxford, Dept. of Astrophysics,
	   Denys Wilkinson Building, Keble Road, GB-Oxford OX1 3RH\\
        \and
           Also at Department of Physics, University of California at Berkeley,
           366 Le Conte Hall, Berkeley, CA 94720-7300\\
           email: smengel@eso.org, mlehnert@mpe.mpg.de, thatte@astro.ox.ac.uk, genzel@mpe.mpg.de}

\date{Received ............... ; accepted ............... }

\maketitle

\begin{abstract}

Accurately determining the star formation history in NGC 4038/4039 --
``The Antennae'' is hampered by variable and sometimes substantial
extinction.  We therefore used near infrared broad- and narrow-band
images obtained with ISAAC at the VLT and with SOFI at the NTT to
determine the recent star formation history in this prototypical merger. In
combination with archival HST data, we determined ages, extinction and
other parameters for single star clusters, and properties of the cluster
population as a whole.

About 70\% of the K$_s$-band detected star clusters with masses
$\geq$10$^5$ M$_{\sun}$ are younger than 10 Myrs (this is approximately
an e-folding time for cluster ages), which we interpret as evidence for
rapid dissolution but not free expansion.

The total mass of K-band selected clusters is about 5 to 10$\times$10$^8$
M$_{\sun}$  and represents about 3-6\% of the total molecular gas.
However, this takes into account only the detected clusters and in view
of the rapid dissolution means that this is only a lower limit to the
total mass of stars produced in clusters during the burst.  Studies of
cluster formation in other galaxies recently suggested short cluster
dissolution timescales, too, which means that star formation rates may
have been severely underestimated in the past.

Extinction is strongly variable and very high in some regions, but
around A$_V$=1.3 mag on average. Even though most clusters are detected
at least in I-band, only the information about individual cluster ages
and extinction allows to avoid uncertainties of orders of magnitude
in star formation rate estimates determined from optical fluxes.
From the distribution of individual cluster extinction vs. age, which is
significantly higher for clusters below 8-9 Myr than for older clusters,
we infer that this is the time by which a typical cluster blows free of
its native dust cocoon.
\end{abstract}

\section{Introduction}
NGC 4038/4039 -- ``the Antennae'' -- is one of the closest (D $\sim$
19.3 Mpc, H$_0$ = 75 kms$^{-1}$Mpc$^{-1}$) examples of merging spiral
galaxies and thus has been the target of detailed studies in many
wavelength ranges \citep[e.g.,][etc]{W99, Fabbiano, Wilson01}, some aimed
at unveilling the star formation properties. However, the term ``unveil''
is appropriate also in a literal sense, since many of the active star
formation sites -- especially in the region where the two galaxies
seem to overlap (therefore often referred to as ``overlap region'')
are hidden by up to several optical magnitudes of extinction.  Mid- and
far infrared wavelengths are essentially not affected by this, but the
spatial resolution of the images is too low to resolve the star formation
regions into single star clusters. We therefore aimed for a compromise
by observing the merger in the near infrared, where the extinction is
only $\sim$10\% of that at visible wavelengths, but where ground-based
telescopes currently deliver sub-arcsecond resolution if atmospheric
conditions are favourable.  We combine the information from these broad
band (Ks, ISAAC on the ESO-VLT) and narrow band (CO 2.34$\mu$m, ISAAC
and Br$\gamma$, SOFI on the ESO-NTT) images with archival Hubble Space
Telescope WFPC2 images \citep[broad-bands roughly corresponding to U,
B, V, I and narrow band H$\alpha$, see][hereafter W99]{W99} for our
investigation of the properties of the numerous compact star clusters
in the Antennae.

What we determined from these data are a number of properties of the
single star clusters, namely  their ages, the extinction towards them,
and their photometric masses.  Additionally, we derived properties of the
ensemble of star clusters as a whole: The age distribution over the 
last $\approx$200 Myrs (including
the mass production in the burst), the average extinction for cluster
populations of different ages, and various other estimates.  The analysis
is complemented by an estimate of the bias introduced by using only the
Ks-band detected clusters, be it an age or mass selection effect.

\section{Observations and data reduction}\label{observations} ISAAC
(VLT-ANTU) imaging of NGC 4038/4039 was performed in ON/OFF mode during
the nights 15.04.2001 (Ks-band) and 16.04.2001 (CO-band-head filter).
The target fit completely onto the detector (0\farcs1484/pixel, total
field size 2\farcm5 $\times$ 2\farcm5).  Seeing was excellent during both
of these photometric nights (PSF FWHM of the co-added frames is below
0\farcs4), and the total on-source time was 360s (Ks) and 480s (CO),
respectively.  No narrow band continuum observations were performed,
therefore the Ks-image was used for continuum subtraction.

\begin{figure*}
\begin{center}
\psfig{figure=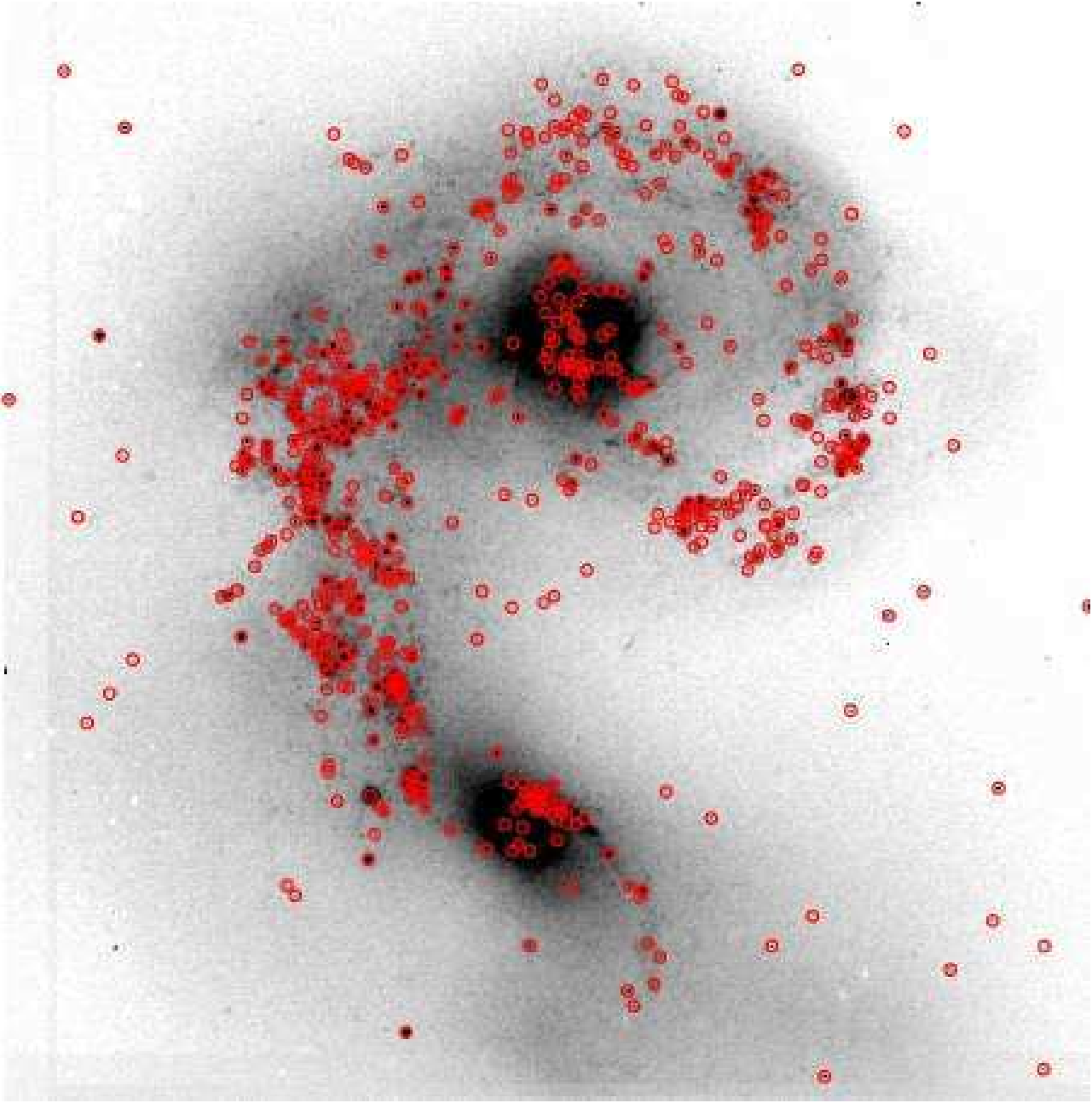,width=17cm}
\caption[Ks-band image, detected clusters marked]{Ks-band ISAAC image
with marked location of the 1072 detected point-like sources.  FWHM of
the PSF in this image is below 0\farcs4. Not all of the detected sources
are clusters, there are also a few foreground stars and background
galaxies. \label{Ksum}}

\end{center}
\end{figure*}

SOFI (NTT) imaging during 11./12.05.99 covered all NIR broad- and several
narrow bands, however, seeing was much worse than for the ISAAC data
(between 0\farcs6 and 1\farcs2 PSF FWHM), so that only the 0\farcs7
Br$\gamma$-data will be used for the analysis. Here the field size was
roughly twice that of ISAAC (4\farcm9 $\times$ 4\farcm9, with a pixel
size of 0\farcs292). Therefore, to save observing time, an on-chip offset
pattern was observed instead of the usual ON/OFF mode. Due to a slight
haze, conditions were not photometric on either of the nights.

Table \ref{obstab} lists descriptions of the observations analyzed in
this paper for both, the ISAAC and the SOFI runs.

\begin{table}
\begin{center}
\begin{tabular}{llllcl}\hline\hline
Band & $\lambda_{cen}$  & Width	& DIT 	& Total ON & Seeing \\
     & [$\mu$m]	& [$\mu$m] & [s]   & [s]	& FWHM	\\\hline
Ks		& 2.16	& 0.27 & 6	& 360	& 0\farcs4\\
Br$\gamma$		& 2.167 & 0.028 & 30 	& 1500 	& 0\farcs7\\
CO$_{2.34}$    	& 2.34 & 0.030 & 60	& 480	& 0\farcs4\\
\end{tabular}
\caption{Summary of the NTT-SOFI and VLT-ISAAC NIR imaging observations.\label{obstab}}
\end{center}
\end{table}

Reduction of the ISAAC data was performed using the IRAF
package\footnote{IRAF is distributed by the National Optical Astronomy
Observatories, which are operated by the Association of Universities
for Research in Astronomy, Inc., under cooperative agreement with the
National Science Foundation.}.  It included dark and sky subtraction
(either using the median of several neighbouring sky images or, where
this led to residuals, doing pairwise subtraction), and flat fielding by
a normalized median of all sky frames. All of the ON frames were slightly
offset with respect to each other, in order to minimize the effect of
pixel defects. Therefore, they had to be shifted to a common location
before using the {\sl imcombine} task (setting the minmax rejection
algorithm to reject the highest and the lowest pixel) to combine the
single frames. A photometric standard (GSPC S279-F, Ks-magnitude 12.031)
was used for flux calibration (the resulting zero-point was 24.28 mag).
The data reduction procedure for the CO narrow band images was identical.

The resulting Ks-band frame is displayed in Figure \ref{Ksum}, with
circles indicating the point-like sources that were detected using {\sl
daofind} from the {\sl DAOPHOT} package.  A detection threshold of
3$\sigma$ was chosen, leading to 1072 detected objects, which are marked
in that Figure.

Reduction of the SOFI data followed essentially the same procedure, with
the two differences 1) that there were no real OFF frames, therefore
several images with observing times closest to the currently treated
frame were medianed for sky subtraction.  2) due to a wrong setting of
the "Pupil Rotation" switch, the object orientation varied from frame
to frame, especially during the transit of the object. This had to be
accounted for before the co-addition of the frames. They were rotated by
the angle derived from the header information.

HST WFPC2 images were reduced using the IRAF {\sl stsdas.hst\_calib.wfpc}
package, and photometric calibration used the zero-points and colour
transformations given in Holtzman et al. (1995) in their table 7
and equation 8.  For H$\alpha$, we set the zero-point to reproduce
the magnitudes of several stars and clusters lacking H$\alpha$
emission (which we know from our 3D \cite{Mengel01} and ISAAC spectroscopy
\cite{Mengel02} data) for a linear interpolation between V and I magnitudes at the
central wavelength of the H$\alpha$ filter (18.316 mag).

\begin{figure}
\psfig{figure=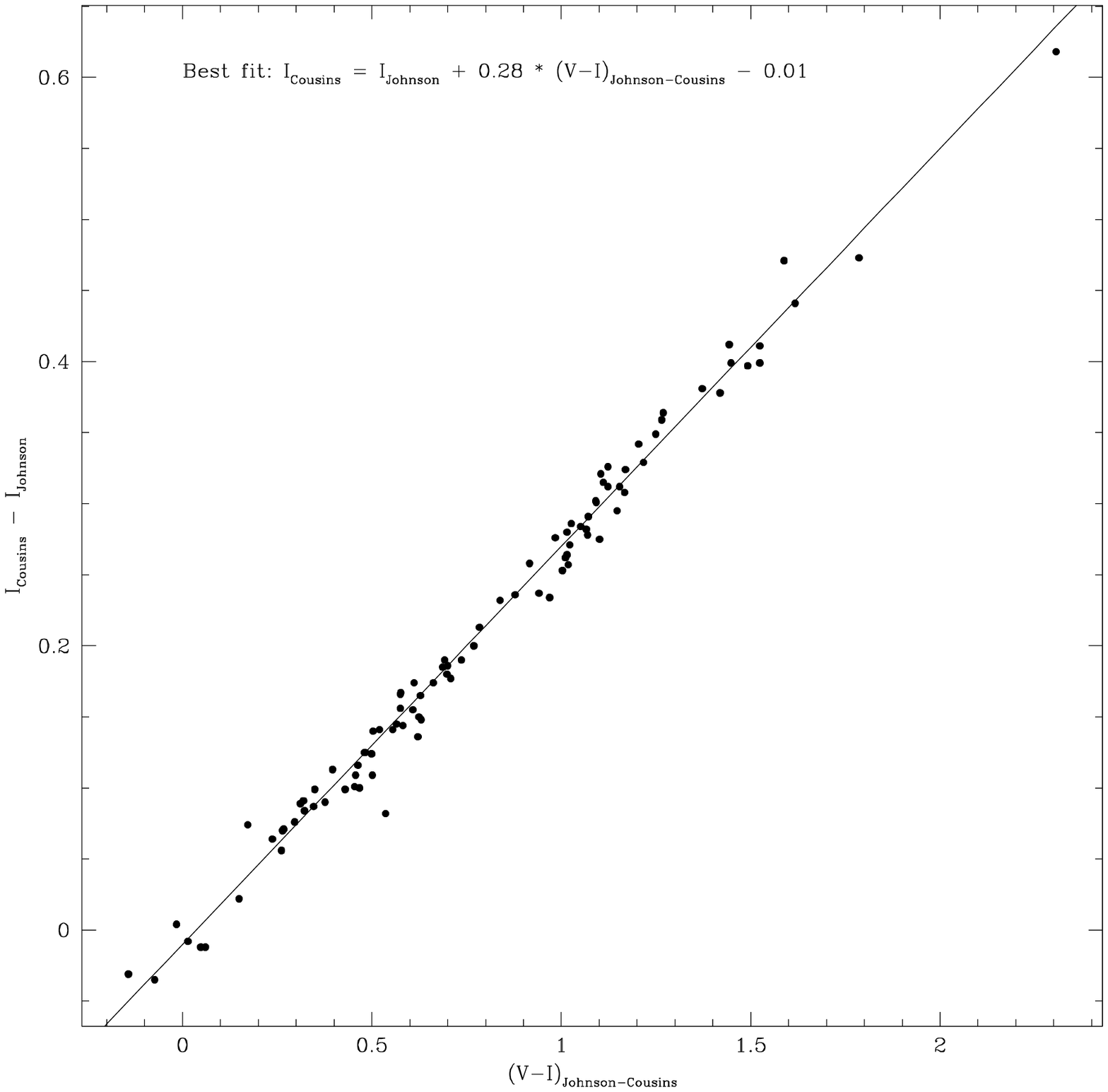,width=8.8cm}
\caption[Transformation Johnson-Cousins for I-band]{ V-I colours in the
Johnson-Cousins system and the difference between the Cousins and the
Johnson I-band magnitudes are plotted for stars of spectral types B2 to
K4, taken from \cite{MoffettBarnes79} and  \cite{Landolt83}. Over-plotted
is the best fit to the data, which is represented by the given analytic
expression.  \label{Ibandcalibration}}

\end{figure}

However, in order to be able to combine the optical and the near-infrared
data sets, we binned the WFPC2 images down to the pixel size of the
ISAAC frame and rotated them to match orientation. Therefore, in the 
following, whenever the unit ``pix'' is used, it refers to the ISAAC
Hawaii array imaging scale of 0\farcs148/pix. 

Photometry was then
performed on the target list which was created using the Ks-band image.
Running {\sl daofind} on the V-band image results in more than 10,000
detected point-like sources. However, the analysis of the HST data alone
was thoroughly performed by W99. Since we are interested in the additional
information brought about by the K-band data, we restrict our analysis
to those objects which are detected in Ks.

In order to estimate the total cluster magnitudes which are required for
the mass estimates of the clusters, and also in order to obtain colours
combining the optical and near infrared bands, we used a curve of growth
technique to estimate the magnitudes of relatively isolated clusters
(distance to nearest cluster more than 10 (12) pixels in V (Ks) - band).
We assigned the magnitude at a radius where the magnitude did not change
significantly any more. For a number of those clusters we determined
aperture corrections for both bands and applied these to the clusters
in the more crowded regions, where cluster distances were below the
mentioned limits. Due to the different PSF FWHMs in the two bands (V:
1.6 pix, Ks: 2.6 pix), we used different fixed apertures (V: r =
3 pix, Ks: r = 5 pix), and the aperture corrections were 0.7 mag and
0.5 mag for V and Ks, respectively. So, in summary, for the isolated
clusters we used the curve-of-growth technique, and therefore variable
apertures, to obtain the total magnitudes, whereas for clusters with
nearby neighbours, we performed aperture photometry and applied an
average aperture correction.

We then determined optical colours (U-B, B-V, V-I) all within the same
apertures (r = 3 pix), since the PSF did not vary significantly over
the optical wavelength range.

For the computation of H$\alpha$ fluxes, we used the magnitudes of
the H$\alpha$ photometry within the r = 3 pix aperture and subtracted
a continuum flux estimated from a linear interpolation of the V- and
I-band fluxes at the central wavelength of the used narrow band filter
($\lambda_{cen}$ = 658 nm). We decided to take such a small aperture to
avoid including flux from neighbouring clusters in the cases of crowded
regions. Radial plots performed on some of the isolated, bright clusters
in the continuum-subtracted H$\alpha$ image showed us that most (70\%-95\%)
of the emission line flux was indeed contained within this aperture.
However, in the crowded regions with many clusters around 7 Myr, 
overlap in the emission line flux is quite common (see the red channel
of Figure 4 in \cite{W99}). The largest shells of some older clusters
have radii of more than 20 pixels, and we clearly cannot integrate out
to these distances without including a lot of neighbouring clusters.
Our selection of a 3 pix radius is therefore the attempt to optimize
the result for young clusters with strong emission. For older clusters
with extended, low-luminosity shells, this will lead to age estimates
which are too old, but we will only use the information from the H$\alpha$ 
emission for strong H$\alpha$ emitters, anyhow.

\begin{figure}
\psfig{figure=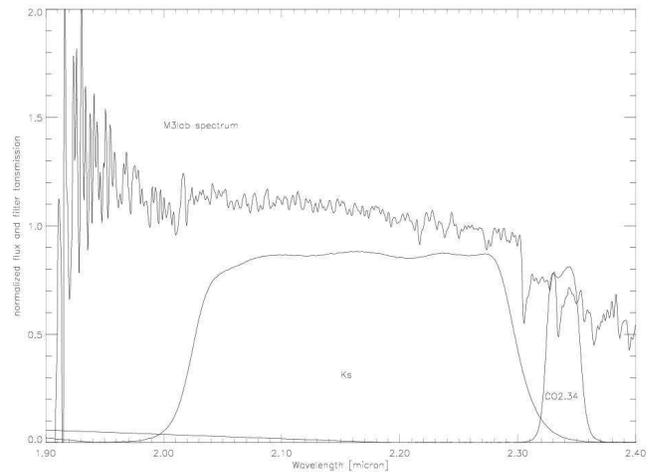,width=8.8cm}
\caption[Definition of CO index]{Two curves which display the transmission
of the CO$_{2.34}$ narrow band filter and of the Ks broadband filter
installed at VLT-ISAAC.  The spectrum is a red supergiant of spectral 
type M3Iab from the stellar library of \cite{Lancon00}, which was shifted 
in wavelength appropriate for the redshift of the Antennae galaxies. 
Note that the Ks filter cuts off just short of
the CO band-head and is not substantially influenced by its absorption.
\label{filtertrans}}

\end{figure} 

\begin{figure}
\psfig{figure=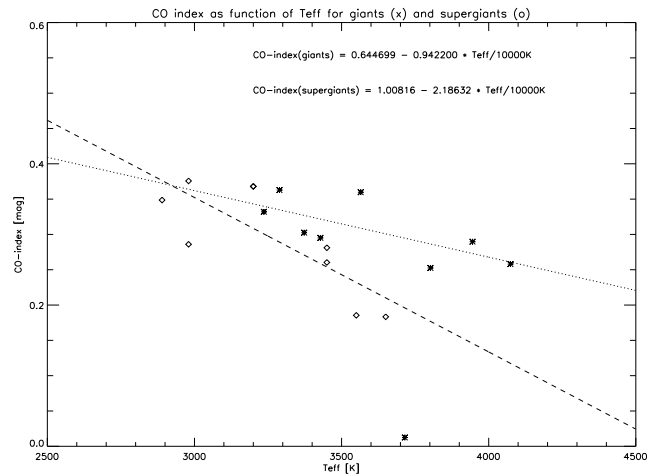,width=8.8cm,angle=90}
\caption[Definition of CO index]{The CO index as a function of effective
temperature estimated for the redshifted spectra from the stellar library
of \cite{Lancon00} using the ESO filter transmission curves from
Fig. \ref{filtertrans}.  The lines are the two best linear fits, for
supergiants and giants, respectively.  \label{co_g_sg}}

\end{figure}

Br$\gamma$-fluxes were more difficult to obtain, due to the much lower
spatial resolution.  Here we proceeded as follows: We convolved the
Ks-band image with a Gaussian, so that the PSF FWHM matched that of
the Br$\gamma$ image. Then we performed photometry on both images (the
zero-point of the Br$\gamma$ image set such that the K-band magnitudes of
stars and clusters known to lack Br$\gamma$ 
emission\footnote{from our 3D \citep{Mengel01} and ISAAC 
spectroscopy \citep{Mengel02} data} were reproduced) for
the 1072 point source positions detected in the higher resolution image.
We then used the Ks-band fluxes as the continuum fluxes, and determined
Br$\gamma$ fluxes and equivalent widths from the difference of the
two images.  In the locations of extremely strong Br$\gamma$ emission
(equivalent width above 150 \AA), the Br$\gamma$ flux is likely to be
underestimated by 5-7\%, due to the line contribution to the continuum
filter.

CO (2.34$\mu$m) images had the same PSF FWHM as the K-band images,
therefore the procedure was facilitated here compared to the SOFI narrow
band data.  We set the zero-point of the image to reproduce the magnitudes
of stars and isolated clusters known to lack CO band-head 
absorption$^2$  
(Z = 21.475 mag). The CO index is created taking the ratio of the flux in the
CO narrow band filter with that in the Ks-band filter and expressing it in
magnitudes. Using the broad-band filter for an estimate of the continuum
should be fine, because it cuts off just short of the rest wavelength of
the first CO band-head (see also Fig. \ref{filtertrans}).  Only in case of
extremely high Br$\gamma$ equivalent widths (above 150 \AA), a CO index
of up to 0.07 mag can be obtained despite the absence of CO absorption.
The aperture used for determination of the CO index was 5 pix.

\section{Analysis and Results}\label{analysis}

We used several different approaches to estimate the ages and extinction
in individual clusters based on broad- and narrow-band photometry (we are only
talking about ``clusters'', because even the brightest supergiants
with M$_K$=-12 mag are unfortunately still below our detection limit).
Subsequently, these methods were combined to yield the cluster properties.
Broad-band photometry was used to simultaneously estimate cluster age
and extinction by fitting the (usually) five data points to colours from
a Starburst99 \citep{Letal99} model for ages up to 500 Myrs.

In order to be able to compare our photometry with the predictions of
Starburst99, we applied three corrections to the models: 

\begin{itemize}

\item In V-band and Ks-band, we accounted for line emission included in the filter
bandpass.

\item The filter systems on board HST do not exactly match the assumed
filter transmission in the Starburst99 code, but differences are
only significant for I-band. We therefore transformed the theoretical
I-magnitudes to the HST photometric system.

\item The CO index computed by Starburst99 is a spectroscopic index
which does not exactly correspond to the index that results from our
choice of filters, so we derived a new analytic expression for the
variation of CO index with effective temperature and inserted this in
the Starburst99 code.

\end{itemize}

\begin{figure}
\psfig{figure=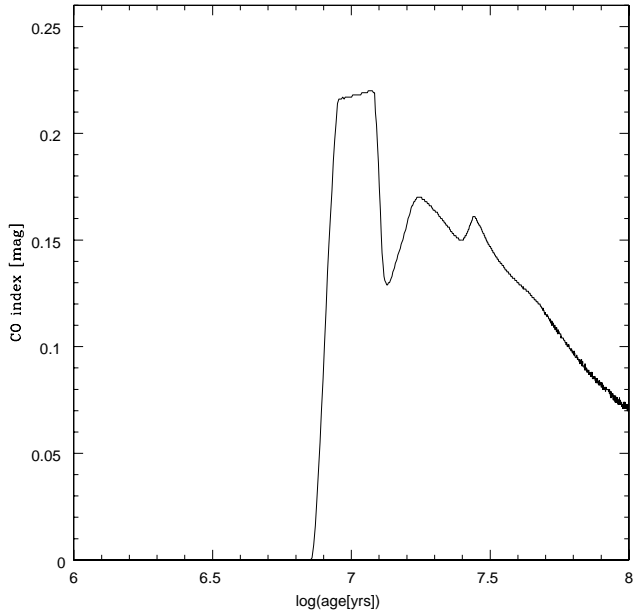,width=8.8cm}
\caption[CO index with ISAAC filters]{The evolution of the CO index using
the ESO CO$_{2.34}$ and Ks filters, as it is predicted by Starburst99
after applying corrections for the definition of the index. The general
behaviour is similar to other CO indices or equivalent widths definitions
with only the absolute values differing.

\label{sb99_COmod}}

\end{figure}

The procedures we followed were in detail:

\subsection{V-band and Ks-band: inclusion of line emission}\label{Vemissionlines} The
HST broadband filter that most closely resembles the Johnson V-band filter
is F555W, and the zero-points and transformations described in section
\ref{observations} actually convert the magnitudes to the Johnson system.
The transmitting wavelength range of this filter includes several emission
lines, amongst them H$\alpha$ and H$\beta$,

We made a crude attempt at taking this line emission into account in
the theoretical predictions of the Starburst99 code.  The H$\beta$
line will contribute the largest amount of hydrogen recombination line
emission, because at the position of the H$\alpha$ line (6563 \AA), the
filter transmission is already down at the $\approx$10\% level, while the
intrinsically fainter H$\beta$ line (4861.33 \AA) lies in the region of
highest transmission ($\approx$90\%).  Also the [OIII] line lies in the
region of high transmission, and even though its strength depends on environmental
parameters, it is typically much stronger than H$\beta$. We included its
contribution by using the ``typical Galactic HII-region'' line ratios 
listed in \cite{AndersFvA03}, amounting to a factor of 5.5. To determine the absolute line
fluxes we used the V-band magnitudes and the H$\alpha$ equivalent widths
predicted by the Starburst99 code, for the strength of the H$\beta$
line we assumed the theoretical line ratio of H$\alpha$/H$\beta$ of
3.3 for Case B recombination \citep{Osterbrock74}. Modifications in
the predictions of V-band magnitude are only noticeable for cluster ages
younger than $\approx$6 Myrs.  In a similar way, Br$\gamma$ emission was
included in the Ks-band filter predictions. However, the contribution
is quite small.  Only for clusters below 4 Myrs, the difference in
Ks-magnitude amounts to more than 0.05 mag.

\begin{figure*}
\begin{minipage}{17.6cm}
\begin{minipage}{8.7cm}
\psfig{figure=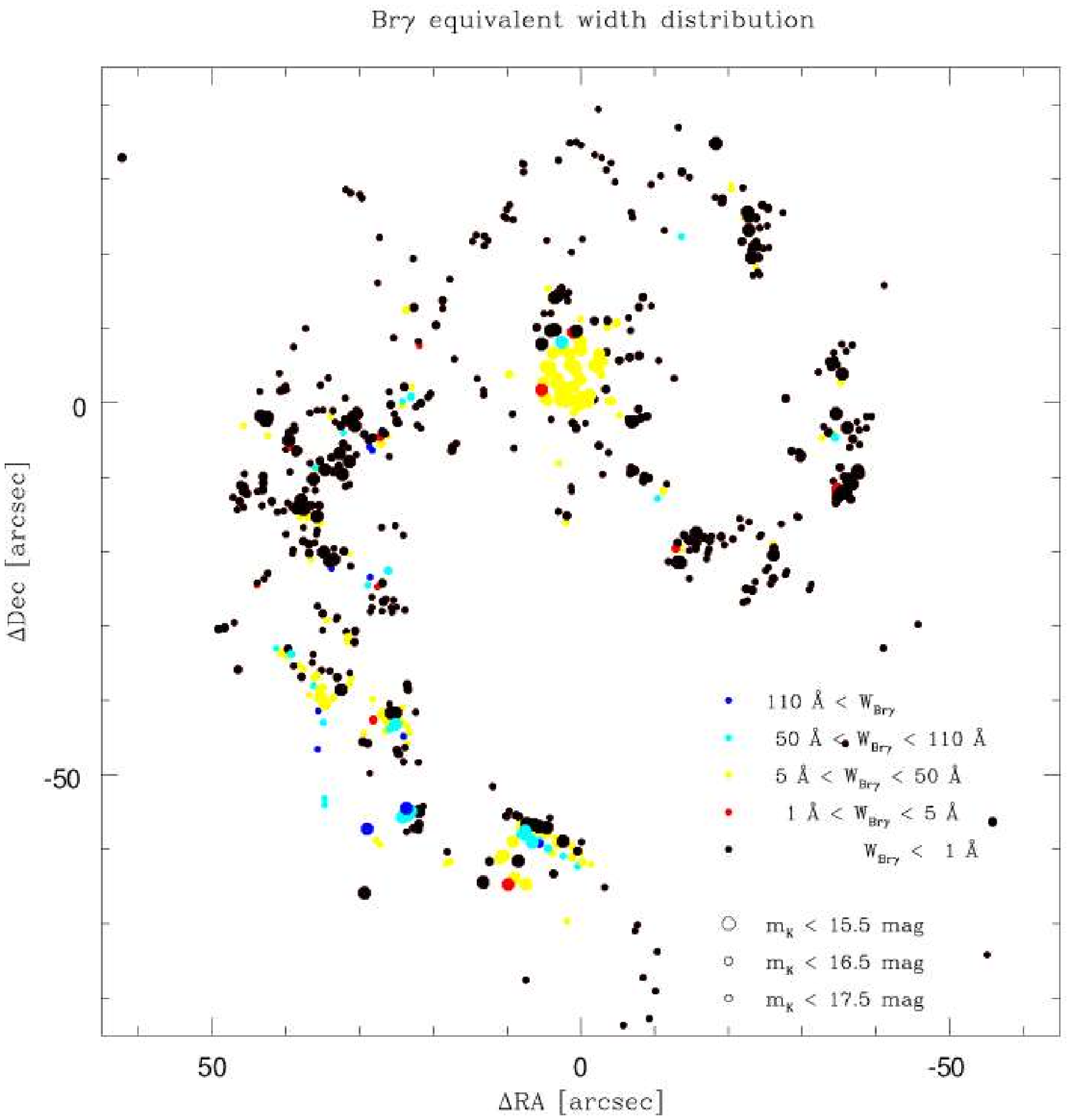,width=8.8cm}
\end{minipage}
\hspace{0.2cm}
\begin{minipage}{8.7cm}
\psfig{figure=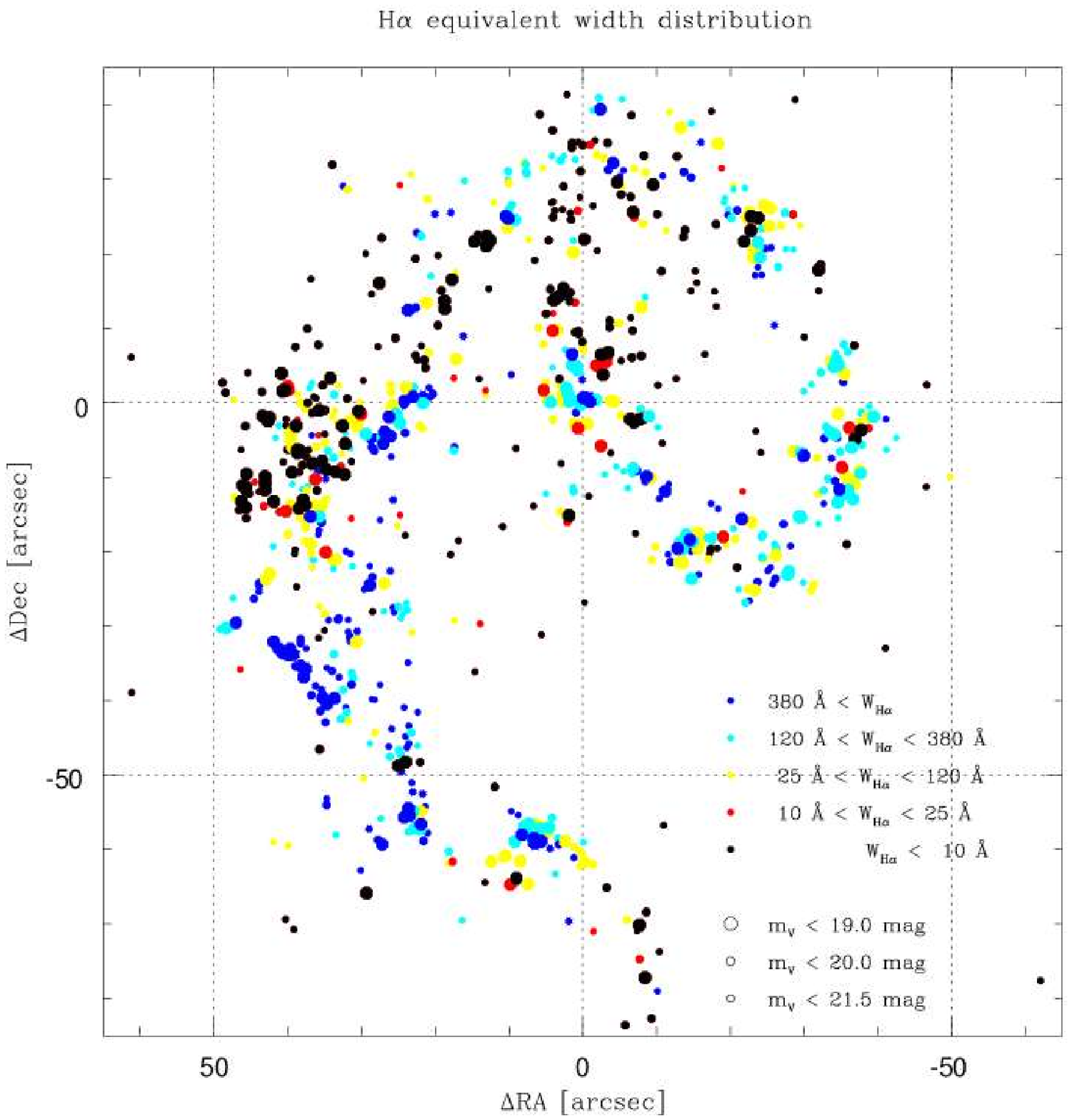,width=8.8cm}
\end{minipage}
\end{minipage}
\vspace{0.2cm}
\begin{minipage}{17.6cm}
\begin{minipage}{8.7cm}
\psfig{figure=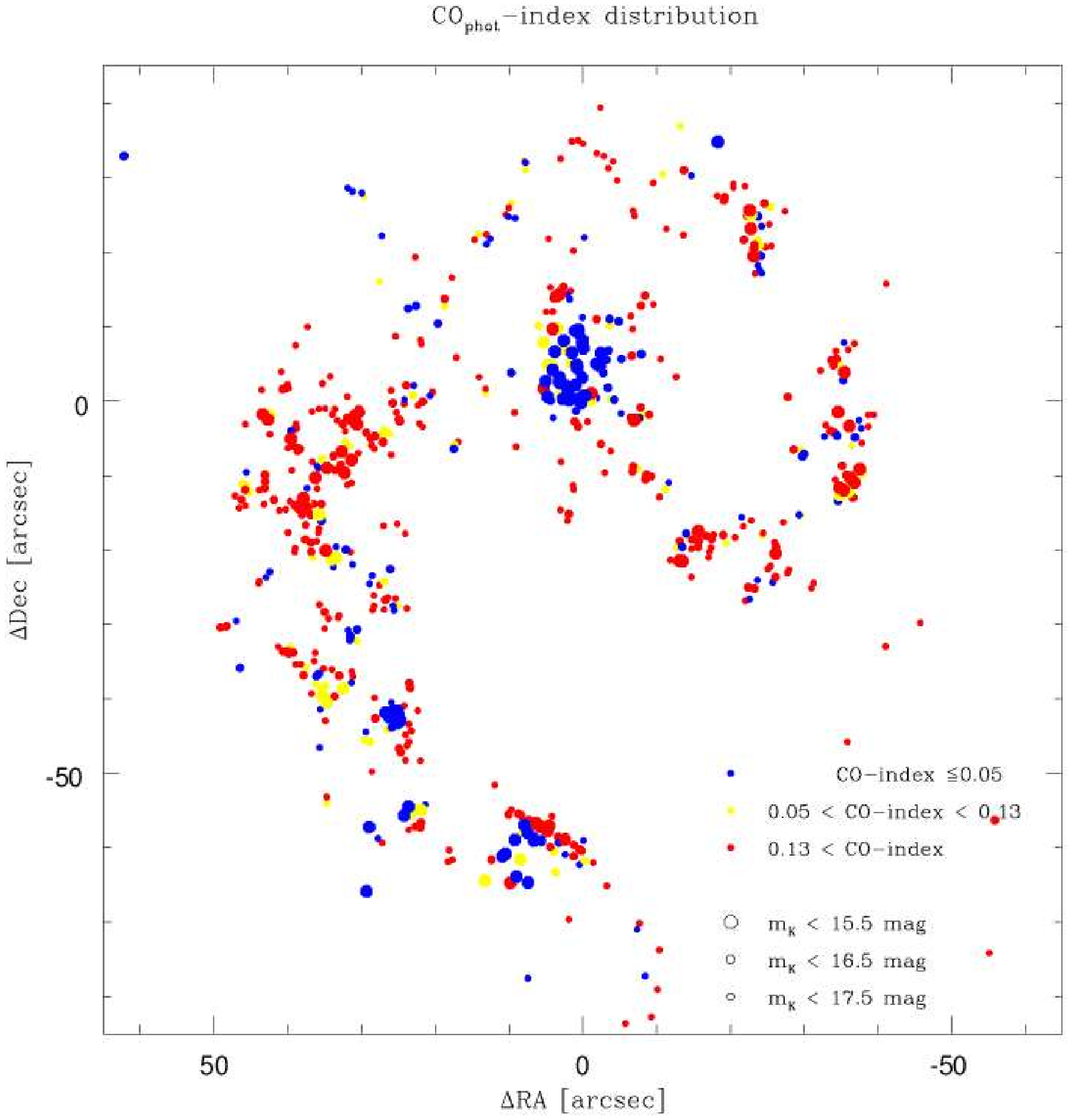,width=8.8cm}
\end{minipage}
\hspace{0.2cm}
\begin{minipage}{8.7cm}
\psfig{figure=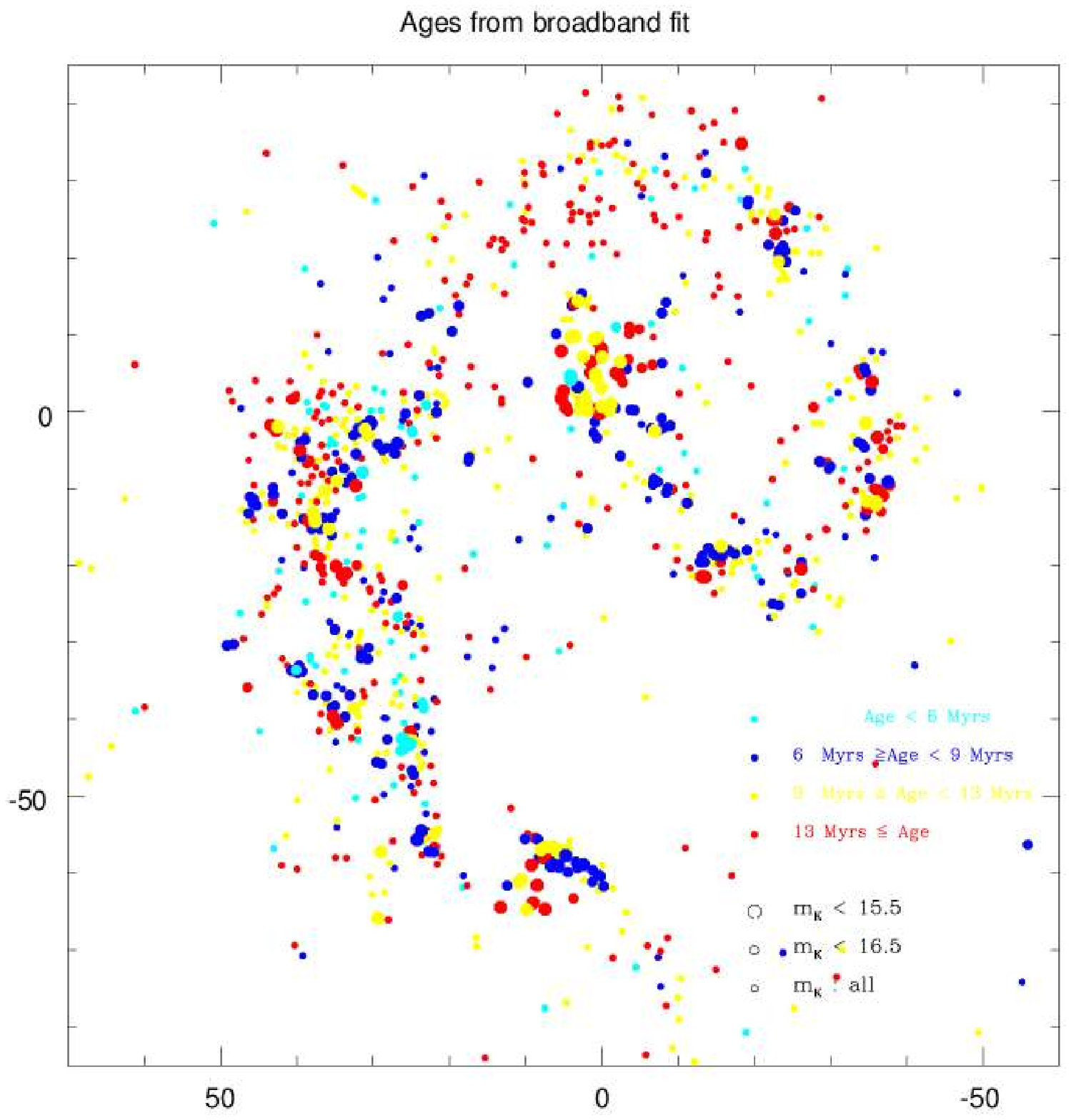,width=8.8cm} 
\end{minipage}
\end{minipage} 
\caption[Comparison: Age distribution from broadband fit and
W$_{Br\gamma}$, W$_{H\alpha}$ and CO index distribution ] {Spatial
distribution of the three narrow band age diagnostics, with the colour
coding as indicated in the individual plots. Generally, blue and cyan
colours indicate younger clusters than magenta and red, but the legends
for each plot should be consulted for individual comparisons. The general
distributions of ages are very similar, and the ages determined from the
various diagnostics coincide within a few Myr for most clusters. There
are a few exceptions, however, where one or more of the indicators show
a deviant value.\label{compareageBrgHaCO}}

\end{figure*}

\subsection{I-band photometry}\label{iphot}

While the HST filter set most closely resembles the
Johnson-Cousins-system, which means that the I-band filter is in
the Cousins system, the Starburst99 code assumed an I-band filter
transmission curve corresponding to the Johnson system. In order to obtain
a transformation between the two systems, we used photometry in both
systems for stars of spectral types B2 to K4, taken from \cite{Landolt83}
for the Cousins system and from \cite{MoffettBarnes79} for the Johnson
system. We converted the Starburst99 I-band magnitudes from the Johnson
system to the Cousins system using the following expression:

\begin{displaymath}
I_{Cousins} = I_{Johnson} + 0.28 * (V-I)_{Johnson-Cousins} - 0.01
\end{displaymath}

Figure \ref{Ibandcalibration} shows the data points we used together with this best fit.

\subsection{CO index}\label{COindex}

As mentioned in section \ref{observations}, we used the narrow band
CO$_{2.34}\mu$m filter and the broadband Ks filter for estimating the
CO index. This arrangement is a close approximation to the photometric
CO index defined in \cite{KH86} who use another narrow band filter for
an estimate of the K-band continuum. The spectroscopic CO index which
is computed in Starburst99 is the spectroscopic CO index defined by
\cite{OO00}.  Rather than attempting a conversion, we used the giant
and supergiant stars in the near infrared stellar library assembled
by \cite{Lancon00}.  The K-band spectra of the stars were redshifted
approximately to the redshift of the Antennae galaxies (z$\approx$0.005)
and we estimated their CO indices using the ESO filter transmission
curves for the Ks and the CO$_{2.34}$ filter.  From this analysis, we
determined two analytic expressions for the CO index as a function of
effective temperature, one for the giant and one for the supergiant stars.
We did not determine the CO indices for dwarf stars, because at cluster
ages below a few hundred million years, their contribution is negligible.
We note that the library of \cite{Lancon00} only included high luminosity
stars, but had we expected a substantial contribution for our purposes,
we would have included them using a different library.

The two analytic expressions we derived are:

\begin{displaymath}
CO index = 0.645 - 0.942 * \frac{T_{eff}}{10000K} \hspace{0.5cm}(giants)
\end{displaymath}
\begin{displaymath}
CO index = 1.008 - 2.186 * \frac{T_{eff}}{10000K} \hspace{0.5cm}(supergiants)
\end{displaymath}

and the data points and the corresponding fits are shown in Fig. \ref{co_g_sg}.

We replaced the corresponding expressions \citep{doyon94} in the
Starburst99 code by these two expressions, then reran it to produce the
evolution of the feature for an instantaneous burst at solar metallicity
for an IMF with Salpeter slope between 1 and 100 M$_{\sun}$.  The result is
shown in Fig. \ref{sb99_COmod}. The general evolution of the feature is
the same as for other age tracers using this spectral feature - like the
above mentioned photometric and spectroscopic indices, or the equivalent
width. Only the absolute values have changed, and they can now be compared
with our observations. See section \ref{ages} for this comparison.

For six clusters, we have information about the equivalent width of the
CO band-head \citep[as defined in][and used in Starburst99]{Origlia98} from
spectroscopic or integral field spectroscopy data \citep[see][]{Mengel01,
Mengel02}. We compare (see Table \ref{COcomptable}) these values and
the corresponding age predictions to the CO index and assigned ages for
two reasons: Firstly, in order to check the reliability of the technique,
and secondly, to assess how strong the influence of extinction on the
observed CO index is. We ended up extinction-correcting the CO index
according to the following formula, which makes a noticeable difference
only for clusters with an A$_V$ of above $\approx$2 mag.

\begin{displaymath}
CO index (extinction corrected)  = CO index + 0.012 * A_V
\end{displaymath}

This accounts for the differential extinction expected from the difference
in central wavelength between the two filters.  The two higher extinction
clusters (W95-355 and W95-80) in Table \ref{COcomptable} would allow
for an even higher correction factor, but we stuck to this physically
justified correction.

\begin{table*}
\begin{center}
\begin{tabular}{llllll}\\\hline\hline
Identification & A$_V$ & W$_{CO}$& CO index (ext.corr) & Age (W$_{CO}$) & Age (CO index) \\
   & [mag] &  [\AA] & [mag] & [Myrs]  & [Myrs] \\\hline
W99-1 & 0.094 & 17.5 $\pm$ 1 & 0.199 & 8.6..12.7 & 8.6..12.5\\
W99-2 & 0.03  & 16.2 $\pm$ 0.2 & 0.165 & 8.5 or 12.8 & 8.4 or 12.7\\
W99-15 & 1.02 & 17.0  $\pm$ 0.2 & 0.279 & 8.6..11.5 & 8.7..12.3\\
W99-16 & $\approx$ 0 & 19 $\pm$ 4 & 0.673$^1$ & 8.2..12.9 & 8.7..12.3\\
WS95-355 & 2.88 & 16.3 $\pm$ 0.2 & 0.134 & 8.5 or 12.8 & 8.2 or 13\\
WS95-80 & 4.01 & 0 & -0.02  &  $\leq$ 7.3 &  $\leq$ 7.3\\\hline
\end{tabular}
\caption{Comparison of ages derived from W$_{CO}$ obtained from spectroscopy
(3D at the AAT and ISAAC at the VLT) with our narrow band determined
CO index. Agreement is generally quite good. For definition of W$_{CO}$,
CO index and extinction correction, please see text.
The identification refers to the numbers used in \cite{W99} and \cite{WS95}, respectively.
$^1$ This value is completely off-scale, even if uncertainties are large. 
The age we assign is the range corresponding 
to the maximum CO index of $\approx$0.22.\label{COcomptable}
}
\end{center}
\end{table*}

As can be seen from Table \ref{COcomptable}, the agreement is quite
good for these clusters, but it is also obvious from the two clusters
with high extinction that it improves agreement to apply an extinction
correction for A$_V$ above $\approx$ 2 mag. It is unfortunate for this
purpose that our spectroscopic studies were heavily biased towards
clusters around 10 Myrs of age, because this narrows down the age range
available for comparison between spectroscopy and narrow band imaging.

\begin{figure}
\psfig{figure=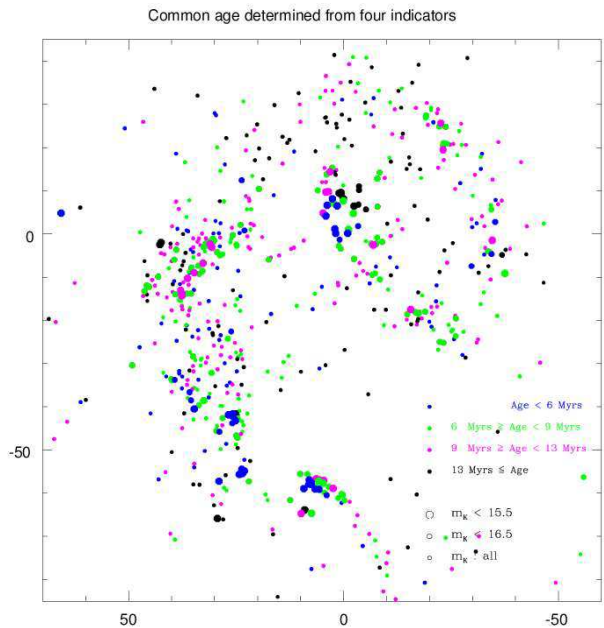,width=8.8cm}
\caption[Clusters with good age determinations]{Clusters with ages
determined from all indicators, only those with good agreement are shown.
The median age is 9.2 Myrs.\label{commonage}}

\end{figure} 

\begin{figure}
\psfig{figure=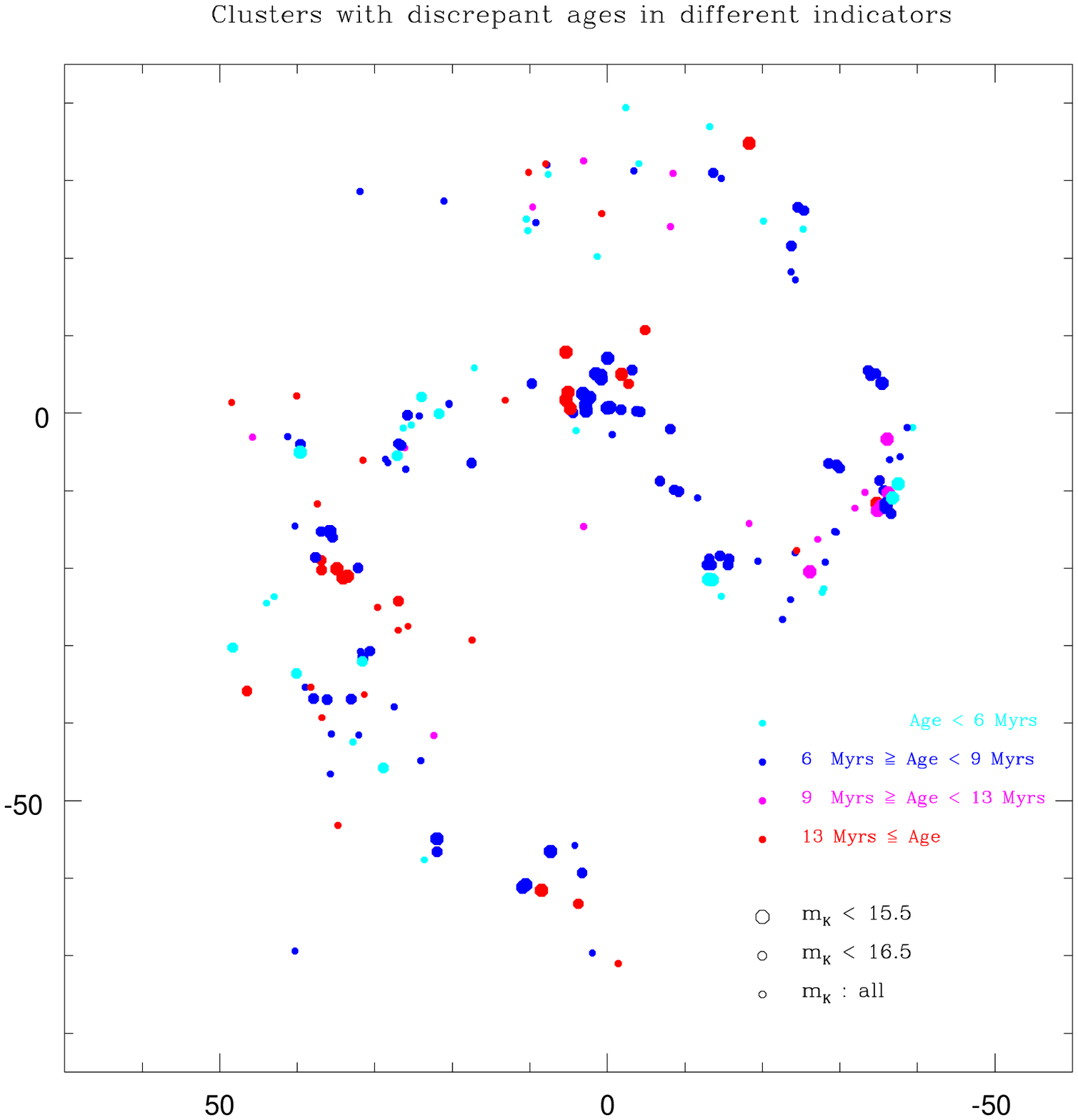,width=8.8cm}
\caption{Colour (age) and size (luminosity) distribution of clusters
that had discrepant age determinations.  Roughly 180 clusters fall into
this category.  The colour coding represents the best age fit determined
from using only broad-band colours. A 2-dimensional Kolmogorov-Smirnov
test indicates that the spatial distribution of these clusters is
marginally different from the distribution of all clusters (probability
of 4.6\%). This is because the clusters that lie in regions of high
background relative to their total brightnesses are more likely to have
discrepant ages than the general population of clusters.  Thus bright
bright clusters in the nuclei, for example, tend to have inconsistent age
determinations, whereas faint clusters with discrepant age determinations
tend to be distributed more like the general population of clusters.
This hypothesis is confirmed by running 2-dimensional K-S tests separately
on the bright and faint sub-populations with discrepant ages compared
to the general population of bright and faint clusters respectively.
The dividing magnitude was chosen to be m$_K$=15.5.
\label{ages_mismatch}}

\end{figure}

\subsection{Our Standard Model}\label{standardmodel}
We used the Starburst99 \citep{Letal99} model, because it focuses on
young stellar populations, which are dominant in this young merger.
Nevertheless, we would have liked a model which also takes into account
the thermally pulsing AGB phase, to be able to assess the impact they have
on K-band luminosity. It is large according to some authors \cite[see
for example][and references therein, and \cite{Schulz02}]{FoersteretalM82}, 
but depends a lot on assumptions like the duration of that phase.  Just before
this paper was ready for submission, the new version of Starburst99
\cite[][including the Padova tracks with full AGB evolution]{VazquezLeitherer05} 
became available. Too late to re-do the
analysis, but in time to add one track to Fig.  \ref{detlimitsVK}. It
shows that the impact on our interpretation would have been marginal if
we had used the new models instead.  The 10 Myr (red supergiant) peak in K-band luminosity
has a smaller amplitude and is somewhat wider in the new model (which
would probably have caused our age distribution to get slightly wider).
We emphasize however that the peak introduced by the population of
TP-AGB stars of around 200 Myr is not pronounced enough to have shifted
a significant fraction of the K-band selected clusters into that age
range within our detection limit. Thus the differences in the models
did not lead to a significant number of incorrect (younger) age assignments.

The characteristics of the model we used throughout this paper (unless
noted otherwise) was an instantaneous burst, with solar metallicity, and
a Salpeter IMF with a low and high mass cut-off of 1 and 100 M$_{\sun}$
respectively. These parameters are justifiable. The instantaneous burst is
applicable because the star clusters are typically very small (effective
radii of a few parsec) and very concentrated, which means that the star
formation event is essentially instantaneous.  

Solar metallicity or just
slightly above is what we determined for a few bright individual clusters in
the Antennae from an analysis of UVES high spectral resolution spectra
(paper in preparation). Note that \citet{VazquezLeitherer05} caution
against using their model (or any current evolutionary synthesis model)
for stellar populations with sub-solar abundances and a significant
population of red supergiants. This is the main reason why we decided to
fix the metallicity at a reasonable average value, rather than making it a
fit parameter, despite the necessary caution suggested by \cite{Anders04}.

The IMF is probably only {\sl on average} around Salpeter
\citep[][]{Mengel02}, and we expect it to vary for the whole cluster
population in a similar manner to what was seen for the subsample
analyzed in \cite{Mengel02}. Since we have not yet found a parameter
which to correlates significantly with IMF slope or lower mass cutoff,
we assume this average value determined by \cite{Mengel02}. Note that
assuming, for example, a lower mass cutoff of 0.1 M$_{\sun}$ for the
Salpeter IMF, or a Kroupa IMF \citep[][]{Kroupa2001} instead, does not
change our conclusions derived from relative quantities like colours
or equivalent widths, because those IMFs largely add mass in the form
of low-mass stars, which do not contribute a significant amount of
light at young ages.  Only the cluster mass derived from the photometry
would be higher than the ones we obtain, therefore our masses represent
lower limits.  For example, the mass of a cluster for a Salpeter IMF
between 0.1 and 100 M$_{\sun}$ is higher by a factor 2.6.

\subsection{Age Determination}\label{ages}

Our data set includes a number of age indicators, therefore we used the
following strategy to assign individual ages to the Ks-band detected
clusters. We deemed this necessary (rather than a simple averaging of
all available ages) because of the on/off nature of the narrow band
diagnostics: For example, absence of CO absorption in a sufficiently
bright cluster constrains the age to be anything below 6.3 Myrs, which
can be used to verify ages from other diagnostics, but not for straight
averaging.  In essence, what we did was:

\begin{itemize} \item Firstly, we have the broadband data covering U, B,
V, I, and Ks, and in principle suitable to age-date clusters of all ages
(the large wavelength coverage, especially the inclusion of Ks-band
photometry, is suitable for breaking the age-extinction-degeneracy -
see also \cite{Anders04} for a more general discussion).

\item Secondly, we can use narrow band emission line images: H$\alpha$
and Br$\gamma$.  The presence of hydrogen recombination line emission
indicates a very young age of the cluster, below $\approx$7 Myrs, and we
use the equivalent width to determine the age if it is below this limit.

\item Thirdly, CO band-head absorption narrow band images, where the
detected spectral feature reveals the presence of red supergiants or
giants, and is most efficient in identifying star clusters around 10
Myrs of age.

\end{itemize}

As mentioned, we used the broadband data to break the degeneracy in
age and extinction.  To do that, we performed a simultaneous fit of the
theoretical spectra for ages up to 500 Myrs to the observed broadband
magnitudes. The amount of extinction was determined by reddening the
theoretical spectrum according to the Mathis (1990) extinction law.
Note that we are assuming a foreground screen extinction. We believe this
to be appropriate, because after the onset of stellar winds and supernovae,
remnant gas and dust are expected to be driven out of the cluster 
within less than a crossing time \citep[see][]{BoilyKroupa03},
which for the clusters we have studied is $\la$10$^5$yr.  This material
forms a cocoon around the cluster which can be described as a foreground
screen. A mixed model would have to be assumed if we analyzed the average
extinction in the galaxy from integrated emission, but not in determining
the extinction towards individual clusters.  To simultaneously determine
age and extinction, we used a simplex algorithm to tour data space,
with a $\chi^2$ minimization technique to find the best fitting pair
of extinction/age.

Ages and extinction so obtained were complemented for the brighter
clusters (only brighter clusters because the narrow band images
were shallower) by extinction estimates derived from the ratio of
H$\alpha$/Br$\gamma$ flux compared to the value expected for a Case
B recombination scenario, and by age estimates from the Br$\gamma$
equivalent width , the H$\alpha$ equivalent width and the CO index.

The average ages were usually assigned by averaging the available data,
but we kept track of the number of available data points and their level
of agreement by assigning an ``age quality'' parameter, which is highest
if all four indicators (broadband fit, W$_{H\alpha}$, W$_{Br\gamma}$,
CO index) are present and agree, and is lower for fewer data points
or disagreement:

\begin{figure}
\begin{center}
\begin{minipage}{8.8cm}
\psfig{figure=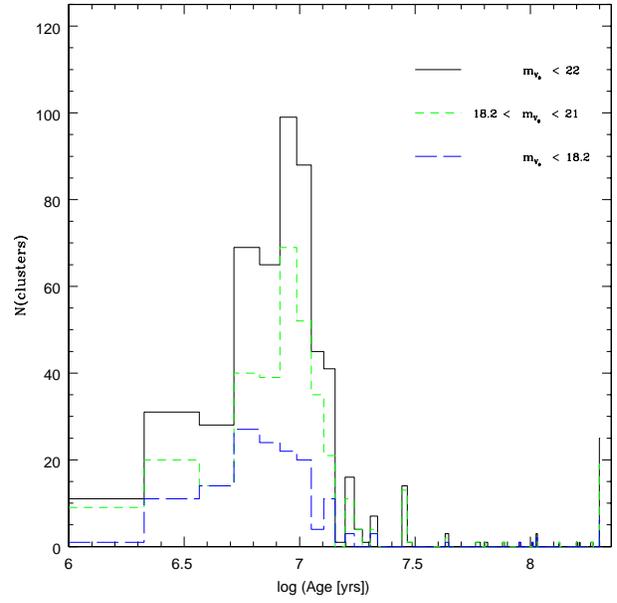,width=8.6cm}
\end{minipage}
\begin{minipage}{8.8cm}
\psfig{figure=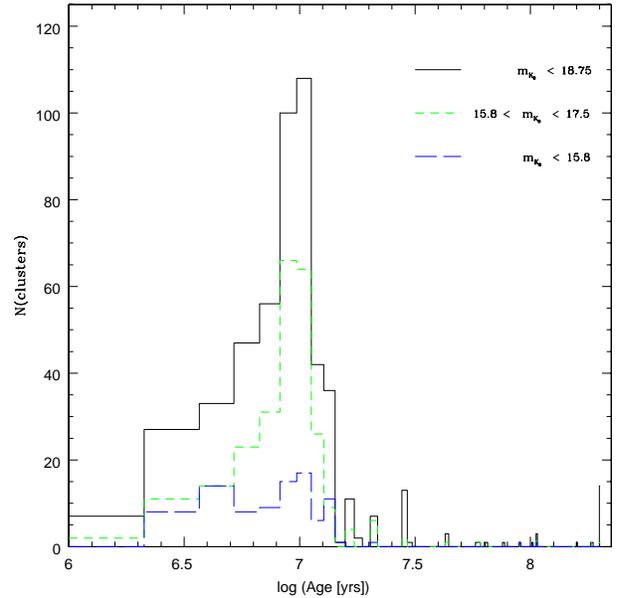,width=8.6cm}
\end{minipage}

\end{center}
\caption{Histogram of numbers of star clusters observed in each age bin.
All clusters are young, and most clusters are either around 10 Myrs or
around 6 Myrs old. Different colours/line-types differentiate between
V-selected brightness limits (top) and Ks-selected (bottom).  The strong
peak of clusters of age 10 Myrs is a selection effect introduced by
detecting clusters in Ks-band, which has a strong peak in absolute
luminosity around that age, caused by high-mass stars entering the
red supergiant phase. This means that many low-mass clusters
are detected in that age bin which would be below the detection limit
at other ages (see also Fig. \ref{detlimitsVK}).  Note the increasing
relative number of young clusters ($\approx$4 Myrs) when going to brighter
clusters in {\sl both} figures, which is already an indication that more
clusters are present at younger ages, an even more convincing evidence
for this being displayed in Fig. \ref{highmassvsage}.\label{agehistogram}}

\end{figure}

\begin{itemize}
\item If all four indicators yielded ages (74 cases), the common age
was the average of the four, if the age difference between any of them
was smaller than 4 Mys. The ``age quality'' was set to 4.  If only three
indicators agreed, their average was used and the fourth value ignored,
with the quality value set to 3.  If two or more ages were discrepant
with respect to the others, the common age was set to the average of
the H$\alpha$- and the Br$\gamma$-age, but the quality-value was set to
a low value of 2.

\item Similar strategies were followed for average age determinations
if fewer than four indicators were available (three: 376, two: 320,
one: 262), essentially averaging the available ages and assigning
data quality parameters depending on the level of agreement. What should
be mentioned is that for the cases where only one age was available
(that would usually be the value from the broadband fit), that age was
also assumed for the average age, and that cases where only one or two
indicators were available, but where the other indicators are expected to
be missing were also assigned high age quality parameters. For example,
a cluster with an age determined from the broadband fit and the CO index
to be older than 10 Myrs is not expected to show H$\alpha$ or Br$\gamma$.
Therefore, it is assigned an age quality parameter of 3.9.

\end{itemize}

\begin{figure}

\psfig{figure=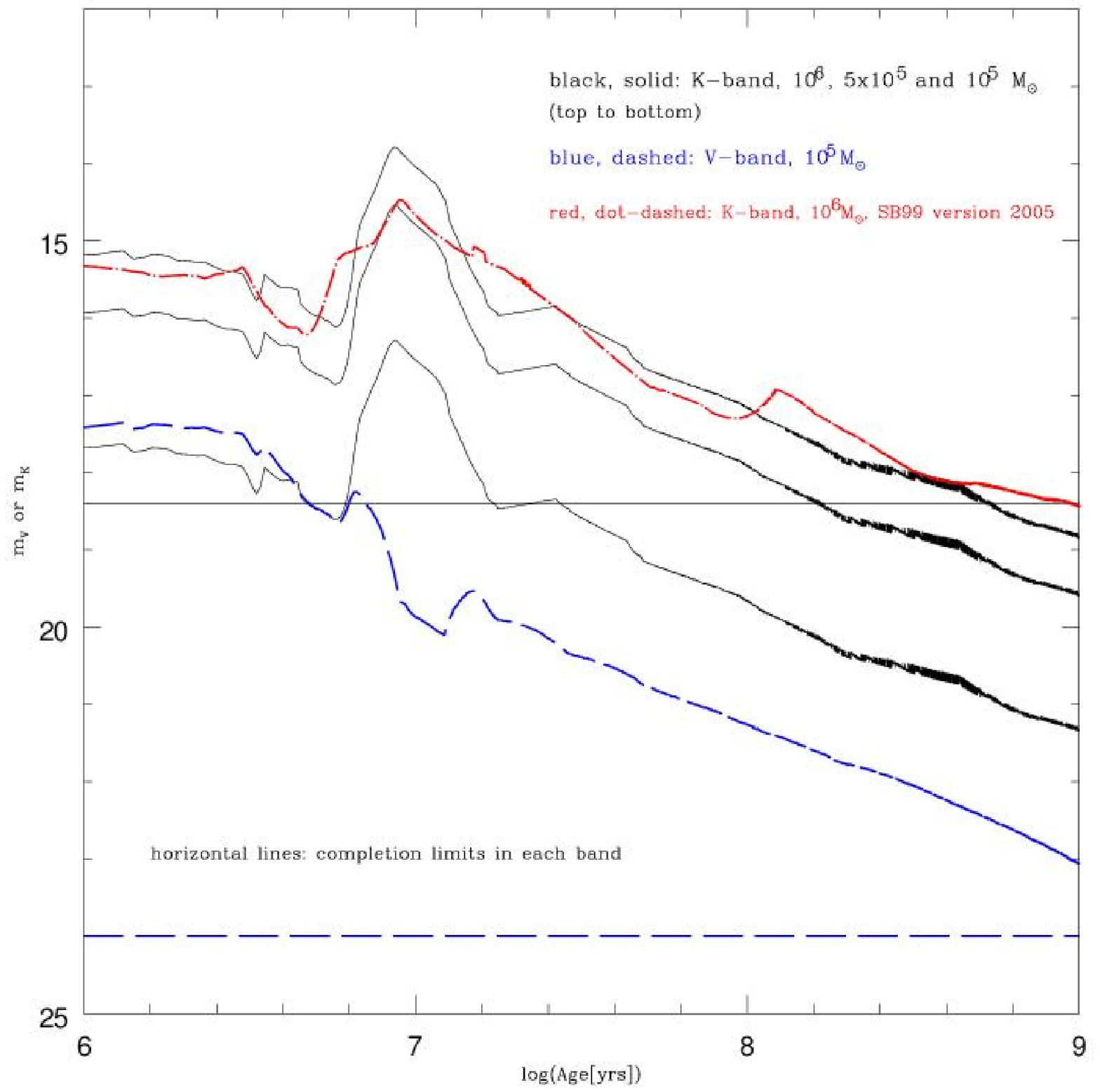,width=8.8cm}
\psfig{figure=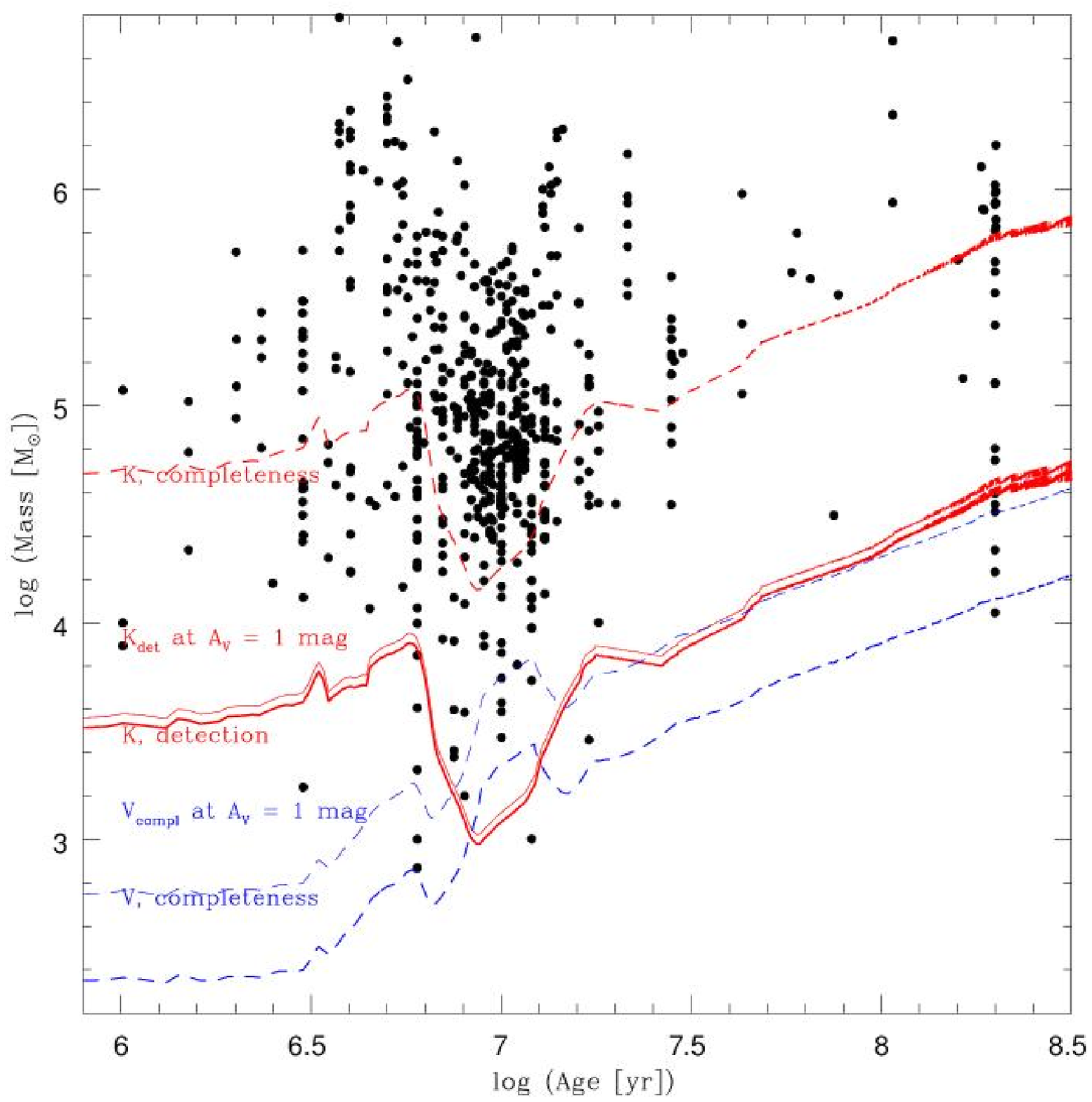,width=8.8cm}
\caption{Two different representations showing the completeness and/or
detection limits for V and Ks band for our clusters.  The upper graph
shows completion limits for clusters with different total masses in
V- and Ks-band (both lower curves are for a 1$\times 10^5$M$_{\sun}$
cluster, for Ks-band, we added 5$\times 10^5$M$_{\sun}$ and 1$\times
10^6$M$_{\sun}$). This shows that in V-band, 50\% of all clusters above
a mass of 1$\times 10^5$M$_{\sun}$ can be detected up to an age above 1
Gyr, whereas this cluster would drop below the Ks-band completion limit
already at 25 Myrs of age.  The red (dot-dashed) line was, like the other
graphs, created using a Starburst99 \citep{Letal99} model (see text for our
standard assumptions and modifications), but this time for v5.0 \citep{VazquezLeitherer05},
which, amongst other modifications, includes the TP-AGB phase, which
is responsible for the bump after 10$^8$ years.  The lower graph is in
support of our claim that many more clusters exist with ages below 25
Myrs, and that this is not a simple selection effect, because it shows
that clusters above a certain mass (but for comparison needs to be above
the completeness- or detection limit for all ages then) are much more
numerous below 25 Myrs than above that age.  In Fig. \ref{highmassvsage},
this claim becomes even more obvious.  \label{detlimitsVK}}

\end{figure}

\begin{figure}
\psfig{figure=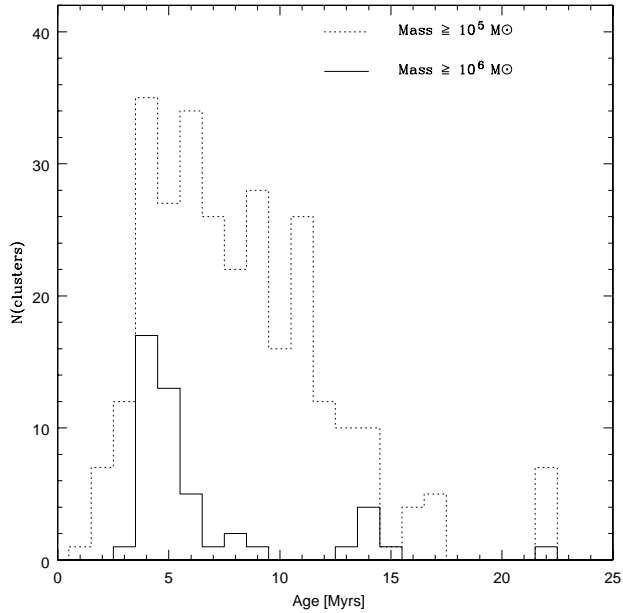,width=8.8cm}
\caption{This plot shows the distribution of clusters above certain mass
limits (10$^5$ and 10$^6$ M$_{\sun}$, respectively), as a function of
age up to an age of 25 Myrs, which is the age where a cluster of 10$^5$
M$_{\sun}$ reaches the Ks-band completeness limit. This means that in
our plot, essentially all clusters above the given mass limits should be
detected, if they are present.  The strong decline in cluster numbers up
to an age of 15 Myrs therefore means that indeed many more clusters are
present around 6 Myrs, meaning that either more clusters were produced
or that a large fraction has been destroyed on timescales of 10-15
Myrs.\label{highmassvsage}}

\end{figure}

Out of our 1072 Ks-band detected clusters, 40 could not be assigned
any ages, 997 had ages from the broadband fit, 590 from the CO index,
128 from W$_{Br\gamma}$ and 494 from W$_{H\alpha}$.  In general, the
agreement between the age indicators was good: excellent agreement
was seen for $\approx$ 340 clusters, another $\approx$ 500 showed good
agreement, and only $\approx$ 230 were vastly discrepant  (which includes
the 40 clusters without an age determination). Their distribution is
shown in Fig. \ref{ages_mismatch}.  Most of the brighter ones of these
objects with bad age determinations were either foreground stars, or located in
the nuclear regions, which suggests that for the latter the high and
variable background of the nucleus might be responsible for photometric
errors that led to wrong age assignments - for example because line
emission was erroneously detected, or subtracted from the source because
it is too extended.  The fainter objects were distributed throughout
the merger, which means that a disagreement in age determinations for
them is probably caused by statistical uncertainties in the photometry.
For  the following analyses, clusters with an ``age quality'' parameter
below 3.0 were not taken into account (which means that foreground stars
with their bad age fit are automatically excluded from the analysis).

The age distribution (see Fig.~\ref{commonage} and
Fig.~\ref{agehistogram}) shows two peaks, one at approximately 5-6 Myrs,
and one around 10 Myrs. Older clusters are rare, there are two more peaks
at around 28 and 200 Myrs, with roughly 20 members each.  The strong
peak around 10 Myrs is a selection effect which arises from the very high
total Ks-band luminosity around that age, leading to detection of large
numbers of low mass (between 10$^3$ and a few times 10$^4$ M$_{\sun}$)
clusters which are below the detection limit at other ages.

However, one interesting aspect of the age distribution is that the
relative numbers of 6 vs. 10 Myr old clusters shifts with limiting
magnitude: Going to brighter clusters, the younger ones become more
dominant.  This is expected from the luminosity evolution in V-band,
where the youngest clusters are the brightest (see Figs. 47b and
51b in \cite{Letal99} or Fig \ref{detlimitsVK} in this work for the
luminosity evolution in V and K), but the absolute Ks-band magnitude
peaks at a cluster age of $\approx$10 Myrs, where the luminosity
contribution from the supergiants dominates that band. Therefore,
we expected to see this trend in the plot of the V-band selected
clusters (Fig. \ref{agehistogram}, top), but were surprised to see this,
even though to a lesser extent, also for the Ks-band selected clusters
(Fig. \ref{agehistogram}, bottom).  Since we used the extinction corrected
magnitudes, it is not extinction which causes this effect. It rather
means that actually more high luminosity (and therefore high mass)
clusters are present at ages around 6 Myrs, compared to those around
10 Myrs.  This conclusion is also supported by Fig. \ref{highmassvsage},
which clearly shows the drop in cluster numbers between 4 and 15 Myrs for
the high mass clusters. We performed a one-dimensional Kolmogorov-Smirnov
(K-S) test in order to assess the significance of the decrease in cluster
numbers. For the clusters above 10$^5$M$\odot$, we created many hundreds
of random comparison populations, one assuming a constant probability
in time, the other assuming an exponential probability distribution, both
with the same total number of clusters as observed.  In neither case
did we concretely account for a selection function because as we have
shown earlier, our selection should be robust for clusters with masses
above 10$^5$M$\odot$ and ages $<$30-100 Myrs.  The probability that the
observed population was drawn from a parent population with a constant
distribution of cluster numbers in time is virtually zero (only 3.5\%).
The population is not consistent with being a simple exponential either
-- only a probability of 56\%.  For the clusters above 10$^5$M$\odot$,
the only simple robust conclusion we can draw without conjuring up more
complex models, is that the distribution of cluster ages is consistent
with our visual impression which is the number of clusters is declining
significantly with age.

\subsection{Extinction}\label{extinction}

A similar procedure like for the ages was followed for the determination
of extinction A$_V$ (we are assuming foreground screen extinction, see
section \ref{ages}).  For most clusters, it was determined from broadband
colours, using the fitting routine described in section \ref{ages} for
a Mathis (1990) extinction law, but for those with m$_K <$ 17.3 mag and
W$_{Br\gamma} > 80$\AA, the value determined from the observed flux ratios
of H$\alpha$ and Br$\gamma$ was averaged with the broadband determined
value by averaging the fluxes and converting back to magnitudes.

\begin{figure*}
\begin{center}
\psfig{figure=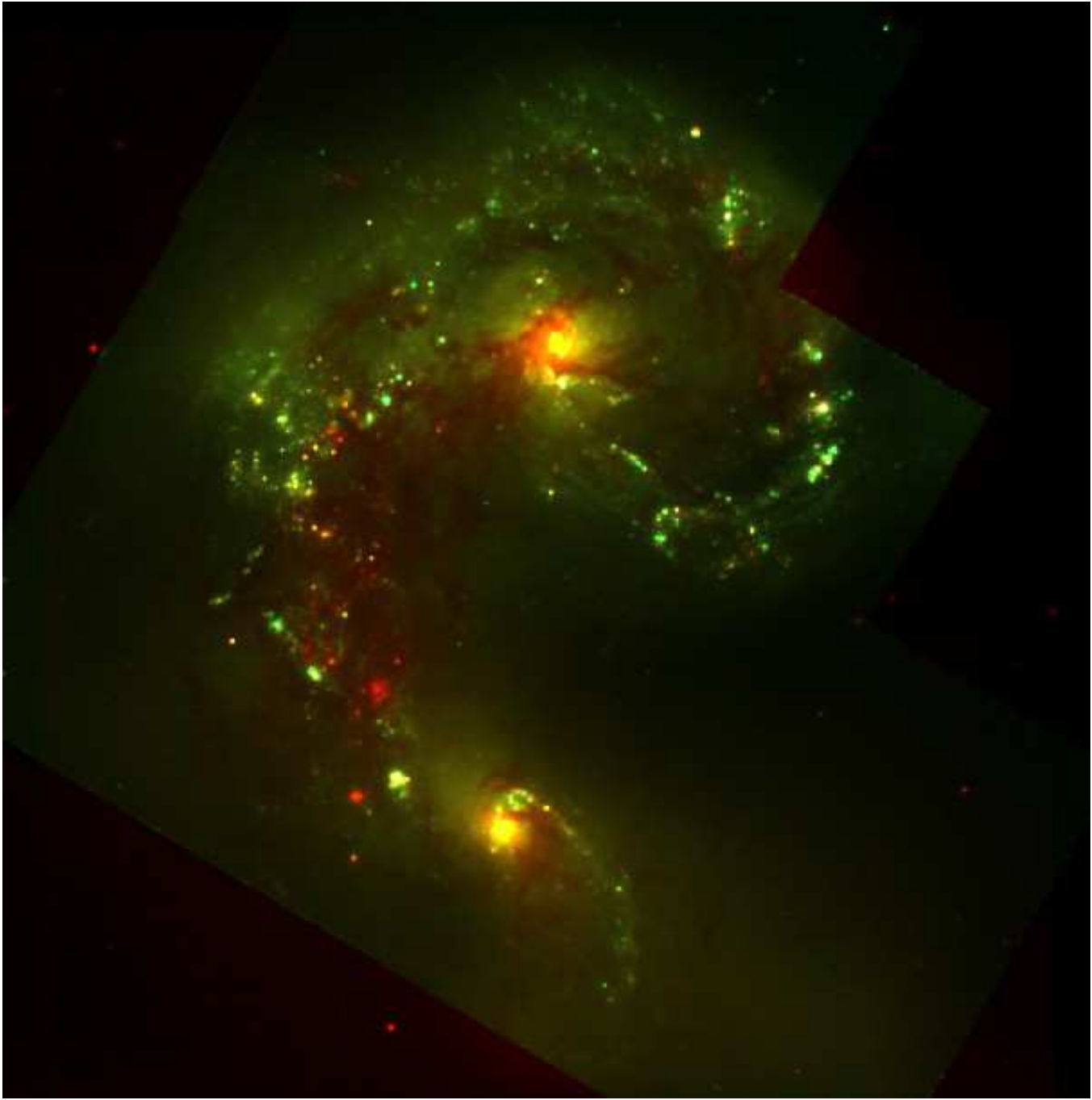,width=17.9cm}
\caption[False colour UBVIKs-image of NGC 4038/39]{\label{UBVIK} False
colour image of NGC 4038/39 with U+B in the blue, V+I in the green (these
four images were obtained from the HST archive, see W99 for details)
and Ks (ISAAC, VLT) in the red channel. Most of the reddening seen
in the clusters in the ``overlap region'' between the two nuclei (and
elsewhere in the merger) is caused by dust absorption and therefore gives
a first-glance impression of the patchy and partially high extinction
encountered in the merger, which was the main driver for obtaining our
near-infrared observations. FWHM of the PSF is 0\farcs4 (HST images were
smoothed), the region displayed covers approximately 2 x 2 arcminutes,
north is up, east to the left. }

\end{center} 
\end{figure*}

As could be expected from just looking at the false colour image
(Fig. \ref{UBVIK}), extinction is very variable within the merger, peaking
at extreme values of A$_V$ above 10 mag in the overlap region. All
extinction values quoted in this article assume a foreground screen
extinction model.

\begin{figure} 
\psfig{figure=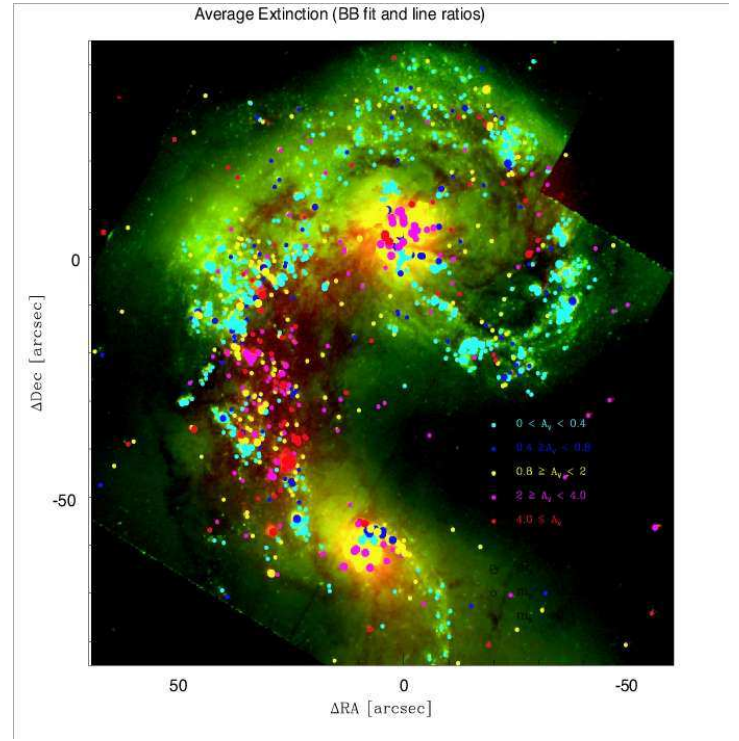,width=9.5cm}
\caption[Extinction distribution]{Spatial distribution of extinction
values (A$_V$) determined for the clusters. Values from the broadband
fit and from the line ratios were combined to a common average where this
was possible. Note the excellent agreement between high extinction values
(red and magenta dots) and the lanes of dust obvious in the multi-colour
image (the same as in Fig. \ref{UBVIK}, only with different cut levels).\label{AVdistr}}

\end{figure}

The distribution of extinction is shown in Fig. \ref{AVdistr}, where we
plotted a colour-coded map of A$_V$ determined from the broadband fit,
and, where possible, in combination with the line ratio value. Underlying
this distribution is the CO map taken from \cite{Wilson01}, to demonstrate
the excellent agreement between high extinction clusters and density
peaks in the CO map.

\begin{figure}
\psfig{figure=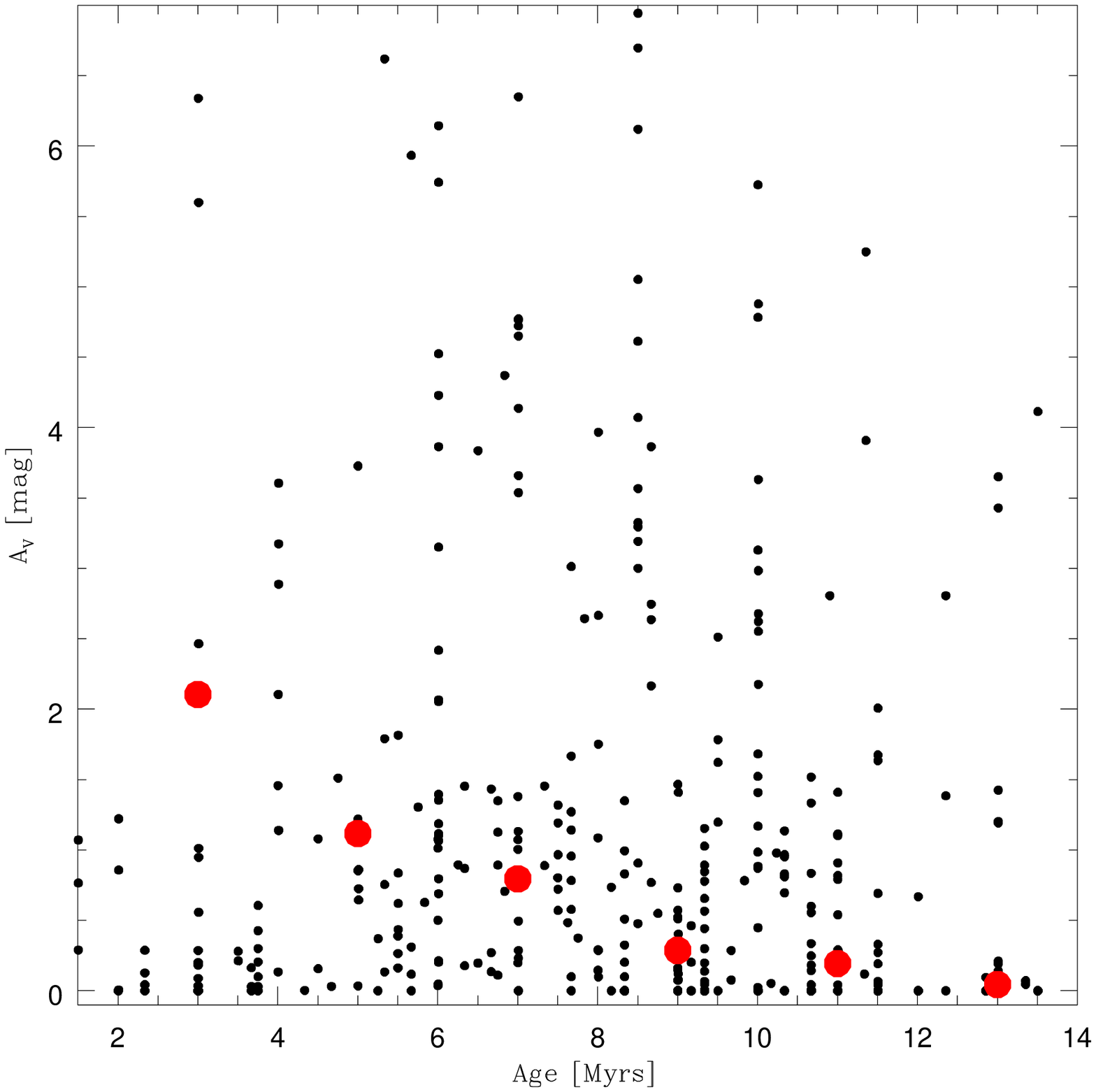,width=8.8cm}
\caption[Extinction-age-relation]{Small black dots are individual
A$_V$, age data points for clusters with good age estimates. The large
red dots are the medians of extinction values found in 2 Myr wide bins
(the last bin shown contains 36 data points, many of which have A$_V$
close to 0 and therefore overlap). Despite the large scatter in A$_V$
values in any given age bin, the expected trend of lower A$_V$ with
increasing age is obvious, and many of the clusters older than 8-9 Myrs
have very low extinction. This lends further support that this is the
age where many of the clusters emerge from their natal dust cocoon.
\label{AV_age_relation}}

\end{figure}

The average extinction is around A$_V$ = 1.3 mag, and it is interesting
to analyze the age dependence of the extinction distribution, reproduced
in Figure \ref{AV_age_relation}, for information about the clearing time
of the clusters.

The typical extinction decreases with increasing age, as can be seen in
Fig. \ref{AV_age_relation}, which shows the median extinction values
as a function of age in bins of 2 Myrs, together with the individual
data points which were used in the median.  See Sec. \ref{conclusions}
for an analysis of this evolution.

\begin{figure}

\psfig{figure=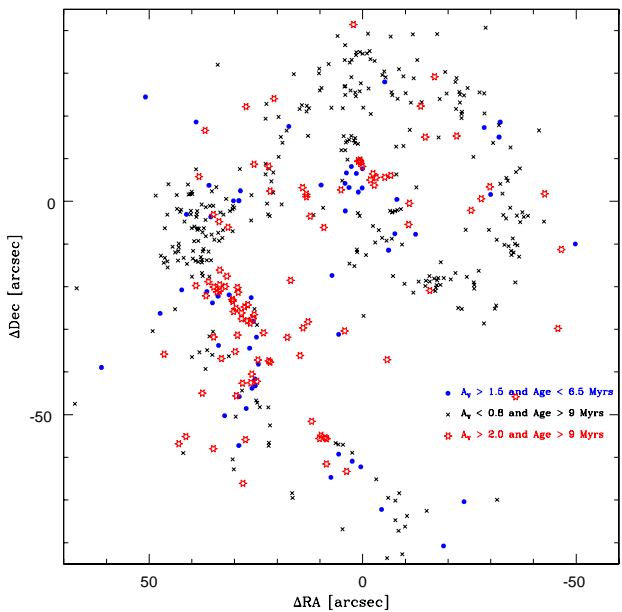,width=8.8cm}
\caption[Extinction-Age-relation]{Extinction values colour coded in their
distribution in the Antennae. The 10 Myr clusters with high extinction
are preferentially located near younger clusters, while those with
low extinction values are often either single or in groups of older
clusters.\label{origin_highAV_10Myrs}}

\end{figure}

These age-averaged numbers artificially increase the extinction values
for older clusters, at least if the purpose of the analysis is to
determine the extinction directly related to the cluster in question,
for example arising from its own dust cocoon. This is suggested by
Fig. \ref{origin_highAV_10Myrs}: Visual inspection of the locations
of high- and low-extinction clusters above 9 Myrs with respect to
the locations of young, high-extinction clusters show that the high
extinction older clusters appear almost exclusively in the direct
vicinity of younger clusters, which we took as an indication that
they only happen to lie within or behind the molecular cloud related
to the formation of the younger, neighbouring cluster. 

In an attempt to quantify the robustness of this visual impression,
we calculated (via a two-dimensional K-S-test) the probability 
that the population of clusters that are
young (age$<$9 Myrs) and highly extinguished (A$_V>$1.5mag), old 
(age$>$9 Myrs) and highly extinguished (A$_V>$2.0mag), or old with 
low extinction  (A$_V<$0.8mag) are spatially distinct.  Our visual
impression is confirmed.  We find that the probability that the young and
highly extinguished and the old and highly extinguished to be spatially distinct
to be very low (only 22\%). Thus it is likely that these clusters are drawn
from the same sample. However this is not the case for the young and
highly extinguished clusters and the old clusters with low extinction.
These samples are have a probability of $<<$1\%  of being drawn from
the same population spatially.

We plotted Fig. \ref{origin_highAV_10Myrs} with lower age limits for the
``older'' clusters increasing from 6 to 12 Myrs in steps of 1 Myr (but we
do not show these plots here), and the described effect becomes obvious
between 8 and 9 Myrs, which we interpret to be the time when most star
clusters have cleared away their natal dust cocoon.  This is the same
age range which is suggested by Fig. \ref{AV_age_relation}.

Mostly caused by extinction, a fraction of the Ks-band detected clusters
is not detected at shorter wavelengths.  To estimate the impact extinction
has on cluster detection and estimates of, for example, star formation
rates based on observations in certain optical bands, we determined which
fractions of clusters were undetected, how much mass they contribute,
and how this compares to the mass which would be derived for the total
cluster mass if only the observed light in clusters for the given band
is used for mass determination.

Table \ref{photcompare} shows that the total fraction of clusters
undetected in any given optical band, compared to the Ks-band, is rather
low, despite the sometimes fairly high extinction, thanks to the fact that
the HST observations are up to a few magnitudes deeper: Even in U-band,
only 16\% of the clusters are not detected at all, and those clusters
contribute only 12\% of the total mass in clusters. These numbers become
even more favourable going to V or I band.

The impact of extinction becomes obvious, however, if one compares the
masses that would be derived from optical cluster light without applying
an extinction correction.  Using the individual cluster ages to derive
individual cluster masses from just the observed magnitudes, like we did
for Ks-band, U and B band lead to a total mass estimate which is just
a quarter of the mass derived from Ks, and this improves only slightly
towards V and I band, where a third of the Ks-band mass is derived.

\begin{table*}\label{photcompare}
\begin{center}
\begin{tabular*}{18cm}{ccccc}
\hline\hline
Band      & number of clusters  detected & mass of clusters not           & mass derived from m$_{Band}$ and \\
          & in Ks, but not in Band       & detected in Band over Ks mass  & age over Ks-mass\\
\hline
 U        & 16\%                         & 12\%                           &  26\%\\
 B        & 13\%                         & 10\%                           &  27\%\\
 V        & 6\%                          & 5\%                            &  33\%\\
 I        & 9\%                          & 6\%                            &  33\%\\
\hline\hline
\end{tabular*}
\caption{This table compares the photometry of detected and undetected
clusters in the optical bands to the Ks band, in order to assess the
impact of extinction on e.g. cluster detection and the star formation
rates derived from them.  The first column gives the names of the
bands (rather, the Johnson equivalents to the HST broadband filters),
the second column gives the number of Ks-detected clusters which are
not detected in the specified band (in percent). The Ks-mass, which
is referred to in columns 3 and 4, was obtained in the following way:
The extinction corrected Ks-magnitudes for individual clusters were
compared to the magnitude expected for a cluster of 1x10$^6$M$_{\sun}$
(using Starburst99 for an instantaneous burst, solar metallicity,
Salpeter IMF between 1 and 100 M$_{\sun}$).  These masses were summed
up to derive the total Ks-band mass.  Column 3 is the total Ks mass
of the clusters not detected in the specified band, whereas column
4 specifies the mass of all the clusters detected in the given band,
if only the not extinction corrected flux in the given band is used
(together with cluster age) to determine the mass.}
\end{center}
\end{table*}

Not taking into account the individual age information, but rather
assuming average ages and extinction can lead to more or less arbitrary
results: For example, assuming an average age of 5 Myrs for the clusters
and converting the observed U-band flux to a total mass yields 33\% of
the mass derived from the extinction corrected Ks-magnitudes (Ks-mass),
whereas assuming an average age of 25 Myrs and an average extinction of
A$_V$ of 1 mag gives 9 times the Ks-mass.

\subsection{Photometric masses}\label{photmass}

Knowing the ages of the clusters is the presupposition for determining
their photometric masses, M$_{\rm phot}$. The extinction corrected Ks-band
magnitudes are compared to those expected for a cluster of a given age and
mass for the model parameters.  We use Ks-band magnitudes, rather than
any of the optical colours for two reasons: Firstly, in Ks-band all of
our clusters are detected, so it is not necessary to extrapolate to one
of the optical colours for an assumed age and extinction, which always
introduces additional uncertainties. And secondly, Ks-band is much less
affected by extinction, which additionally minimizes the uncertainties.

The Starburst99 \citep{Letal99} model parameters we used were again
instantaneous burst, solar metallicity and an IMF with Salpeter slope
between 1 and 100 M$_{\sun}$ (note that for an IMF extending between 0.1
and 100 M$_{\sun}$, the photometric masses would increase by a factor
of 2.6).  From the resulting individual cluster masses we determine the
total mass of stars produced in clusters during this starburst and the
corresponding star formation rate (or rather, lower limits to both, since
there are still many clusters expected to be below the detection limit),
and a cluster mass function which we compare to the luminosity function.

The total mass produced in Ks-band detected clusters with an age
determination is $\approx$430 $\times 10^6$M$_{\sun}$. The error introduced by
not taking into account those clusters without age determination is
fairly small, since it is mostly the faint, low mass clusters which
lack an age estimate. And, as mentioned above, the SFR we determine
here will be a lower limit.  From Fig. \ref{agehistogram}, we see that
the bulk of star clusters are younger than 25 Myrs,
which yields a SFR of SFR$_{clusters} \approx$ 16 M$_{\sun}$ yr$^{-1}$,
which is almost 20 times the Milky Way SFR, and a moderate to normal
value for a starbursting galaxy.  Interestingly, the total molecular
gas mass estimated for the Antennae is approximately 10$^{10}$ M$_{\sun}$
from CO observations \citep{Wilson01} which is substantially greater
than the estimated total cluster mass - approximately a factor of 10-20
(depending on whether the IMF is Salpeter from 1-100 M$_{\sun}$ or a 
reasonable extrapolation like a Kroupa IMF with a minimum mass 
of 0.1 M$_{\sun}$) of the cluster mass reported here.  

Thus the current gas depletion time would be about 250 to 500 Myrs, 
if a star formation efficiency
of 100\% was assumed. For a more reasonable assumption on the star
formation efficiency, like 30\%, the depletion time needs to be scaled
down accordingly, but in any case, this supports further intense
star-formation over the expected merger time scale.

Taking into account our conclusions concerning cluster dissolution,
and instead assuming constant star formation over the last 25 Myr,
our lower limit concerning the star formation rate increases by at least
a factor 2.5, and would then be above 40 M$_{\sun}$ yr$^{-1}$.

\subsection{Cluster Mass Function}

The cluster mass function in merging galaxies is of special importance
with respect to star cluster formation and evolution, because globular
cluster mass functions (and since they have a small range of ages, also
their luminosity functions) are Gaussian (or similarly single-peaked),
with a typical mass around 2$\times  10^5$M$_{\sun}$. However, luminosity
functions of clusters in merging galaxies show a power law with a
typical slope $\alpha = -2$. Cluster mass functions are difficult to
determine, because, like in this work, it requires the determination
of individual ages for these young clusters where the intrinsic
luminosity varies strongly with age. Nevertheless, the mass function
for the Antennae clusters was constructed from HST data by Zhang \&
Fall \citep{ZhangFall}, and found to be a power law with approximately
$\alpha = -2$, with no indication of a turnover at any mass. A turnover
of power laws is consistent with the luminosity function constructed
from the same data by \cite{W99}, with the turnover roughly at 1$\times
10^5$M$_{\sun}$, comparable to the characteristic mass of globular clusters
mentioned above.

From our Ks-band data, we created the luminosity- and mass functions,
and even though the data are shallower in limiting magnitude, we hope
to benefit from the decreased influence of extinction on our data.

\begin{figure}
\psfig{figure=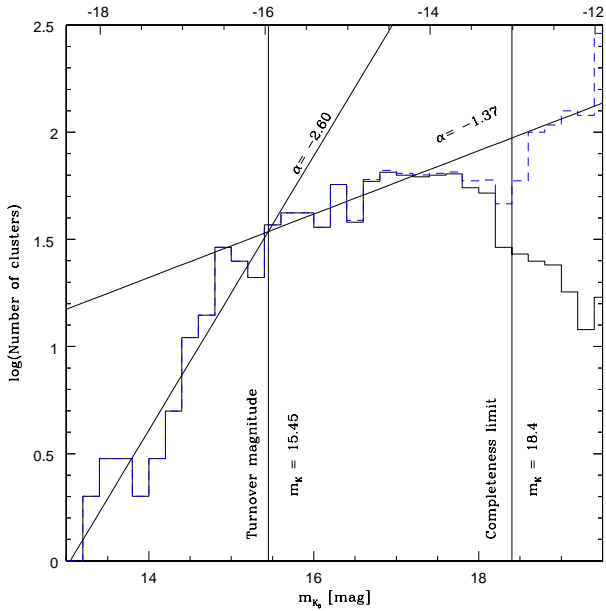,width=8.5cm}
\caption{Extinction corrected Ks-band luminosity function of star clusters
in NGC 4038/4039. Completeness limit (50\% of clusters detected) and
completeness corrected data points are indicated. The data can be fit by
two power laws with a turnover magnitude of m$_{Ks}$ = 15.45.  Comparing
the corresponding absolute Ks magnitude of M$_{Ks}$= -15.96 mag to that
of a 10$^6$M$_{\sun}$ cluster of the median age of 9.2 Myrs (M$_{Ks}$
= -17.52 mag) yields a turnover mass of 2.4$\times  10^5$M$_{\sun}$.
\label{luminosityfunction}}
\end{figure}

\begin{figure}
\psfig{figure=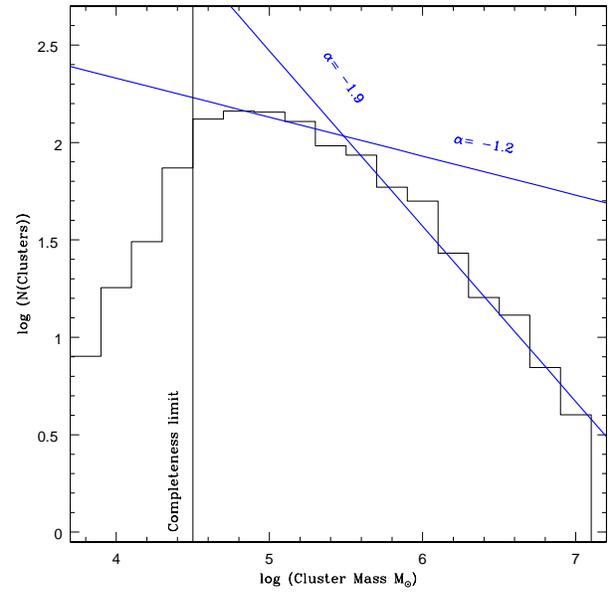,width=8.5cm}
\caption{Mass function of Ks-band detected star clusters in NGC 4038/4039,
and fits using two  power laws (slope -1.9 and -1.2).  Masses were
determined using individual cluster ages and comparing the observed
magnitudes to those predicted by Starburst99 for a 10$^6$M$_{\sun}$
cluster.  The completeness limit is not exact,  because it is the
completeness limit in magnitudes (see Fig. \ref{luminosityfunction})
which was converted to a mass for the median cluster age of 9.2 Myrs.
For some ages which are quite well represented, for example around
6 Myrs, the completeness limit lies rather around 10$^5$M$_{\sun}$,
which makes it really difficult to apply a completeness correction
for this plot, which we therefore omitted. Also these data can be fit
by two power laws, with a turnover mass around 3$\times
10^5$M$_{\sun}$, in reasonable agreement with the number derived from the luminosity
function. The assumption that a large fraction of the highest mass clusters
could be composed of cluster complexes, rather than individual clusters,
could be excluded by visual inspection of both, the K-band image and the higher
resolution I-band image at the locations of the high-mass clusters. At least down to 
the spatial scales which can be resolved, less than 5\% of the clusters with masses
above 10$^5$M$_{\sun}$ look like they might be affected by multiplicity.
However, very close binary clusters (like for example NGC 1569A, which could
be resolved into two components with a separation of 0\farcs2 on HST images \cite[see][]{deMarchi97})
cannot be excluded for the Antennae clusters, which lie at almost 10 times the
distance of NGC 1569.
\label{massfunction}}
\end{figure}

Luminosity and mass functions are displayed in Fig. \ref{luminosityfunction}
and \ref{massfunction}.

The extinction corrected Ks-band luminosity function of star clusters in
NGC 4038/4039 can be fit by two power laws (``broken power law'') with
a turnover magnitude of m$_{Ks}$ = 15.45, and the slopes $\alpha = -2.6$
and $\alpha = -1.37$, respectively. The slopes compare fairly well with
those determined by \cite{W99} from the objects located on the PC chip
which had from an analysis of their Q-parameters been confirmed to be
star clusters.  This fact is somewhat surprising, because  the turnover
masses in the Ks- and V-luminosity functions need not necessarily be the
same, since the cluster luminosity function is a function of cluster ages
(and therefore mass-to-light-ratios) and masses.

Comparing the absolute turnover Ks magnitude of M$_{Ks}$= -15.96 mag to
that of a 10$^6$M$_{\sun}$ cluster of the median age of 9.2 Myrs (M$_{Ks}$
= -17.52 mag) yields a turnover mass of 2.4$\times  10^5$M$_{\sun}$. This
could be interpreted as the progenitor to the typical mass peak in the
globular cluster mass function.

In Fig. \ref{massfunction}, the cluster mass function can also be fit by
a broken power law, with the two slopes of $\alpha = -1.9$ and $\alpha
= -1.2$.  The turnover mass here is around  3$\times  10^5$M$_{\sun}$,
which is in reasonable agreement with the value derived 
from the luminosity function.

However, as we will discuss in Sec. \ref{conclusions}, we do not want to
over-interpret this result, since both random and systematic uncertainties
are rather large, and the latter difficult to overcome.

\subsection{Bolometric luminosity}\label{mbol}

We used the combination of individual photometric masses and ages to
determine the bolometric luminosity expected to be emitted from a given
cluster, using the predictions from Starburst99 \cite{Letal99}, where the bolometric luminosity
is obtained by integrating the spectral energy distribution (without the
nebular continuum). From this, we created an artificial ``bolometric image'' to compare it
to the ISOCAM 15$\mu$m image \cite[][see also the comparison of 15$\mu$m
image and CO map in Wilson et al. 2001]{Mirabel98}, to the radio continuum emission
observed by \cite{NeffUlvestad00}, and to CO emission \citep{Wilson01}.

\begin{figure}
\psfig{figure=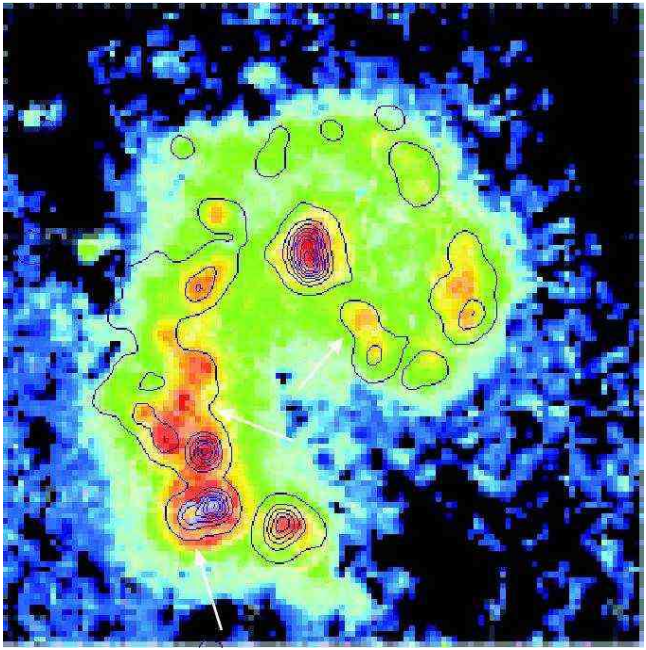,width=8.5cm}
\caption{Overlay of ISOCAM 15$\mu$m image \citep{Mirabel98} with
simulated m$_{bol}$ contours. The simulated m$_{bol}$ image was created
by calculating the expected bolometric magnitude of each Ks-detected
cluster and feeding this, together with the position, into the IRAF {\sl
daophot.addstar} routine. The resulting image was then Gaussian smoothed
to have a resolution comparable to the ISOCAM 15$\mu$m image. In general,
the good agreement between the two maps is noticeable, but apart from that,
we note that the brightest 15$\mu$m cluster (indicated by the arrow at
the bottom) does {\sl not} have the highest bolometric luminosity, and we
identify two other regions where the simulated bolometric luminosity is
substantially lower than expected, compared to the 15$\mu$m flux. Those
three regions are indicated with the arrows. \label{iso_mbol}}
\end{figure}

\begin{figure}
\psfig{figure=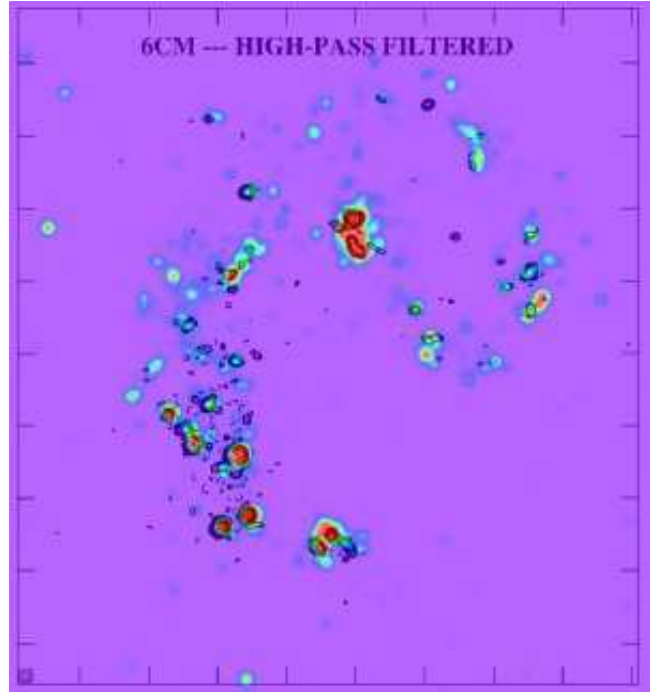,width=8.5cm}
\caption{Overlay of our simulated m$_{bol}$ image (slightly less smoothed than 
in Fig. \ref{CO_mbol})) with radio 6cm (high-pass
filtered, so it shows predominantly the point-like sources) contours
\citep[from][their Fig. 4c]{NeffUlvestad00}. The radio continuum emission here consists
mostly of free-free emission of gas photo-ionized by hot stars, therefore it
is considered a good tracer of recent star formation. Note the excellent 
correspondence between the two maps. \label{mbol_6cm}}
\end{figure}

\begin{figure}
\psfig{figure=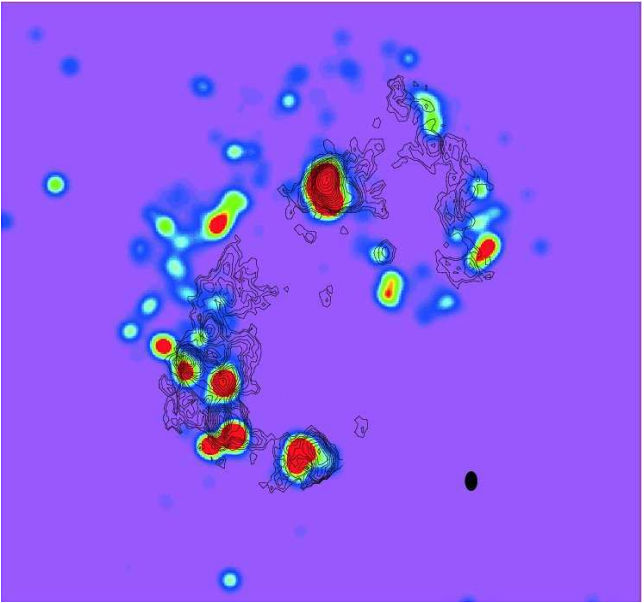,width=8.5cm}
\caption{Overlay of our simulated  m$_{bol}$ image with CO contours
\citep[from][]{Wilson01}. The m$_{bol}$  image is slightly less
smoothed here than in Fig \ref{iso_mbol}.  Rather good agreement is seen
between the two maps in the overlap region, whereas the north-western
region shows a significant offset between the clusters and the CO
peaks. Note that the three regions marked in Fig. \ref{iso_mbol} all lie
in regions where a star cluster lies just on the edge of a relatively
sharp edge in the CO map. The black ellipse shows the beam size of the
CO observations.\label{CO_mbol}}
\end{figure}

Our main goal here was to see whether the mass and age of the very red
star cluster eastward of the southern nucleus (WS95-80) justifies it
being the brightest object in the merger at 15$\mu$m, and how in general
the 15$\mu$m emission relates to the expected bolometric flux. What
we found is that WS95-80 is indeed expected to be a bolometrically
bright object, amongst the brightest sources in the Antennae, but not
as bright as expected from the 15$\mu$m image. While we would expect
it to contribute 3.6\% to the total flux at 15$\mu$m, \cite{Mirabel98}
measured it to contribute roughly 15\%.  Measurements concerning the
contribution of the overlap region as a whole are more in agreement,
while they measure $\approx$ 50\%, we would expect 43\% of the bolometric
luminosity to originate in that $\approx$ 3kpc x 5kpc large region.

\section{Discussion}\label{conclusions}

\subsection{Age Distribution of the Young Clusters}

As shown in Section \ref{ages}, based on Figs. \ref{agehistogram},
\ref{detlimitsVK} and \ref{highmassvsage}, about 70\% of the K$_s$-band
detected star clusters with masses $\geq$10$^5$ M$_{\sun}$ are younger
than 10 Myrs.  Thus, 10 Myrs is approximately an e-folding time for ages
of the most massive clusters.  Unfortunately, our information about
lower mass clusters is severely biased by the form and sensitivity of
mass-to-light ratio in the K-band with age.

There are two plausible hypotheses for explaining this rather narrow
range of ages of the Antennae clusters: Either that at least 70\%
of the clusters have been produced within the last 10 Myrs, or that
the e-folding survival time for the ensemble of clusters is about
10 Myrs and that either the clusters are born unbound or some process
efficiently destroys what are otherwise bound clusters at birth.  We note
again that we can only firmly make this statement for clusters above
$\approx$10$^5$M$_{\sun}$, which lie above the Ks-completeness limit
out to ages of around 25 Myrs. But even considering incompleteness,
the number density of older clusters below our completeness limit still
decline and thus it is probably safe to extend the age limit for
determining the number of clusters out to an age of $\approx$ 100 Myrs.

\subsubsection{Cluster Age Distribution as Star-Formation History}

The changing gravitational potential in a strong merger like the
Antennae can induce strong variations in the star-formation rate
\citep[e.g.,][]{mihos93}.  According to the most common models, closest
passage of the two galaxies comprising the Antennae happened roughly 200
Myrs ago \citep[see e.g.,][]{Barnes88}. At first passage pericenter,
galaxy merger models predict strong star-formation which declines and
depending on the exact geometry and gas distribution, weaker peaks
in the star-formation can occur \citep[e.g.,][]{mihos93, barnes04}.
This enhanced star-formation tends to persist for a factor of 0.5 to
1 dynamical times (a ``dynamical time'' being the orbital time-scale
of the first encounter, thus around 200-300 Myr)
after the first passage pericenter.  However, none
of the models appear to show intense star-bursts that return to the
star-formation rates before the strong interaction occurred.  The contrast
between peaks in the star-formation and the low points during the merger
range about a factor of a few and always well above the rate before the
strong interaction occurred.  In addition, none of the peaks modelled
appear to last less than 5\% of dynamical time.

The e-folding timescale of 10 Myrs which we estimate for the clusters
currently observed in the Antennae is short compared to the dynamical
time of the Antennae.  In the model of \cite{mihos93}, the periapse of
the Antennae occurred about 200 Myrs ago and the galaxies will merge in
about 100 Myrs.  Thus the e-folding time of the cluster ages represents
something like less than 5\% of the dynamical time of the merger.
This is short.  The disks of NGC 4039/39 should have hardly felt a
change in the potential. Nevertheless, clusters with identical ages are
distributed throughout the face of the system, which is hard to explain
as a ``coordinated'' burst, because the communication time between
the regions is larger than the age spread.

It is possible that the interaction has produced other sources
of perturbations like spiral density waves in the disks or bars.
In principal these could also act to induce star-formation.  But again,
the dynamical time scale for spiral density waves is similar to the
dynamical time of the disks themselves which are of the order of 100
Myrs or longer.  Bars can have shorter time scales (10s of Myrs), but
would lead to star-formation which is largely circum-nuclear.  One of
the striking aspects of the star-formation in the Antennae is that
it occurs throughout the disks.  Moreover, we do not observe a large
range of ages across the face of the Antennae as might be expected by
gravitational perturbations.  Since the timescales are long, we would
expect the age range to be a significant fraction of these timescales.
While we do observe that the clusters are younger in the ``overlap''
region, this is only at the level of a few Myrs.  Given the disparities
in the various dynamical timescales, the relatively short e-folding time
of the cluster ages, and the narrow range of ages across the Antennae, it
appears unlikely that the narrow range of ages reflects the star-formation
history within the Antennae.

\subsubsection{Cluster Age Distribution as Cluster Destruction
Time-scale}

On the other hand, there are mechanisms which could lead to the
destruction of clusters on a variety of time scales.  The possible
mechanisms include (roughly in order of increasing time scale): clusters
being born un- or marginally bound, rapid loss of natal material, mass
loss through stellar evolution, stellar ejection during 2-body relaxation,
clusters being born with a significant fraction of their binding energy in
``hard'' binaries, tidal interaction with nearby clusters or gravitational
perturbations such as bars or spiral arms, and gravitational shocks. Of
course, rapid destruction as constrained by the clusters in the Antennae
may require several of these to act in unison.

The simplest hypothesis to explain the short timescale during
which the clusters must be disrupted would be that they are not
gravitationally bound when they form.  If true, this would present a very
interesting quandary given that they are very dense stellar systems.
A model for star-formation has been proposed where the driving force
is supersonic turbulent convergence flows in the interstellar medium
\citep[see the excellent review by][and references therein]{maclow04}.
A situation like that, as opposed to gravitational collapse with 
fragmentation, may lead to compact, vigorous star-formation
as observed in the young compact clusters in the Antennae, but with the
stars that formed not bound as a single entity.

In \cite{Mengel02} we estimated that the crossing time for massive
clusters in the Antennae of $\tau_{CR}=R_{1/2}/\sigma\approx few
\times10^5$ yrs.  The crossing  for less massive clusters than studied in
\cite{Mengel02} will be roughly similar since they have roughly similar
mean densities.  Following \cite{fall04}, if a cluster is approximately
freely expanding as a results of catastrophic mass loss or being born
unbound, the characteristic radius increases as $R_{1/2}(t)\approx
R_{1/2}(0)(\tau/\tau_{CR})$ and its characteristic surface density
decreases as $\Sigma(t)\approx \Sigma(0)(\tau/\tau_{CR})^{-2}$.  Thus
after 10-100 cross times, the characteristic radius has increased by
that amount, and the characteristic surface density would be a factor
of 100 to 10000 lower (5-20 magnitudes).  The size of these effects
means that clusters would drop below our detection limit in only a few
crossing times.

The magnitude of this effect means that it can be tested observationally.
In \cite{Mengel02} we argued from an analysis of the compactness, velocity
dispersion, and IMF of some of the most massive young ($\lesssim$10 Myrs)
clusters in the Antennae that the clusters must be relatively long lived
(survive for more than several 100 Myrs).  This conclusion was reached
under the assumption that the clusters are initially bound and then
subsequent stellar evolutionary and dynamical processes are the only
effects that lead to the dissolution of the clusters.  This study was also
limited to a small number of clusters and certainly could have been biased
to those that are long lived (the ages of the clusters studied were all
close to 10 Myrs).   Our results presented here allow us to address this
question with greater statistical power.  If the cluster were unbound or
dissolving due to internal mass loss (stellar evolutionary and dynamical),
we would expect the cluster to expand and its concentration to go down
with age.  As argued previously, if the clusters were initially unbound
and freely expanding the concentration would decrease rapidly with time.
In an attempt to determine whether the cluster size or concentration
changes with age, which might be an indication of cluster disruption
if older clusters are generally more expanded or shallower than younger
clusters, we measured cluster sizes and King concentration parameters
for clusters in two of our age bins for the K-detected clusters: younger
than 4 Myrs (7 members measured) and between 8 and 11 Myrs (11 members
measured).  In principle, we would have liked more and narrower age bins
and checked for an evolution of size with age, but there are not enough
I-band bright, isolated clusters to fill these bins in statistically relevant
numbers, therefore we used these two clearly separate age bins.

Our result was that the spread in cluster sizes
and concentrations is large for both groups (in the results, we quote
the standard deviation), but that the average sizes are smaller for
the older group, and they tend to be more concentrated. Young clusters:
R$_{eff}$ = 16 pc ($\pm$15pc), King concentrations 15-40, older clusters:
R$_{eff}$ = 6.5 pc ($\pm$5.3pc), King concentrations 7-27. If the clusters
were simply freely expanding, the effect should be strong and obvious.
Thus free expansion seems unlikely.  These results do not lead to a clear
interpretation for other processes that might lead to rapid dissolution
and whose effects are less obvious.  Either, cluster evolution generally
concentrates the clusters, or the weakly concentrated clusters
get disrupted, which decreases the size average for the survivors.

For clusters which are not very concentrated after formation and/or
had a shallow IMF, relaxation processes can lead to destruction
on timescales of a few tens of Myrs \citep[see,
e.g.,][]{TPZ,Mengel02}.  Relaxation processes could lead to rapid
cluster destruction if the cluster is only marginally bound, as might
be the case say for a strong convergence flow like formation mechanism
\cite{maclow04}.  Given there is no clear dependence with age on any
of the parameters measured, we have no unambiguous answer what process
is dominant.   Processes that lead to rapid dissolution short of free
expansion are stellar mass loss (winds and supernova) that reduce both
the stellar mass and remove the natal material (in conjunction with the
radiation pressure), and relaxation processes (e.g., 2-body relaxation,
interactions with hard binaries) that preferentially eject the low
mass stars.  Dissolution of clusters is enhanced for clusters with
shallow IMFs and low stellar concentrations.  So a plausible explanation
for the generally rapid dissolution of a significant fraction of the
clusters is that some are marginally bound and thus the processes such
as loss of the natal material, stellar evolutionary-driven mass loss,
and dynamical mass loss all become more effective and more efficient in
unbinding some of the clusters.

If rapid dissolution of young compact clusters\footnote{The first reference
we are aware of of a dissolution timescale of $\approx$10 Myr was made by
\cite{LadaLada} for Galactic open clusters and therefore not strictly
relevant for our compact, massive targets}
 --- which is also a
favoured scenario for Fall \cite[see][which also focuses on the clusters
in the Antennae]{fall04} and for Bastian et al. \cite[][which analyzes the
cluster population in M51]{Bastian05} --- is common to all
merging galaxies, then estimates of the fraction of star-formation
in mergers (and perhaps other types of star-bursts) that occurs in
such clusters is likely to be severely underestimated.  For example, 
\cite{Meurer95} estimated that about 20\% of the UV light from starbursts
comes from young compact star clusters.  This would suggest that either
there is background star-formation that does not reach the high stellar
densities of the young compact star-clusters or that the background light
is composed of stars from dissolved clusters.  Our results indicate that
this is likely the case.  Indeed, \cite{tremonti01} in a UV spectroscopic
study of cluster and background populations in the dwarf galaxy NGC 5253
found a relative paucity of O-stars in the background population
compared to the compact clusters.  They interpreted this difference
as the background light (non-cluster light) being due to clusters that
dissolved on the order of 10 Myrs.  Combined with our direct statistical
evidence for cluster destruction \citep[see also][]{ZhangFall, fall04}
provides powerful evidence for both a short lifetime of clusters and that
cluster formation must be a very significant mode of star-formation in
all star-burst galaxies.

\subsection{Nature of brightest 15$\mu$m peak}

With its observed age of 4 Myrs and a photometric mass of just above
7x10$^6$M$_{\sun}$, cluster WS95-80 is expected to have a bolometric
observed magnitude of m$_{bol}$=11.4 mag.  This is far less than what is
expected from the 15$\mu$m image, where it contributes 15\% of the total
flux from the merger \citep{Mirabel98}.  This discrepancy is in support of
a suggestion made by \cite{Wilson01} for the origin of the strong 15$\mu$m
emission: One of their two options was that the giant molecular clouds,
three of which seem to be overlapping or colliding in this location,
efficiently transport very small dust grains into the vicinity of the hot
stars which are present in the relatively young (4 Myrs) star cluster.
There, they are heated up to $\approx$100 K, which leads to strong
continuum emission around 15$\mu$m (the spectrum is observed to be rather
steeply rising in that region). For their other option, an even younger
star cluster (younger than 1 Myr), we see no evidence, even in Ks-band or,
from visual inspection, in the 8$\mu$m Spitzer image presented by
\cite{Wangetal04}.

We identify two other regions where we might expect a strongly rising
continuum in the 15$\mu$m range: Roughly in the middle of the overlap
region, and to the south-west of the northern nucleus (marked with
arrows in Fig. \ref{iso_mbol}). Both of these locations, like cluster
WS95-80, have a 15$\mu$m flux which is substantially brighter than the
one expected from the artificial bolometric image.  We did not analyze
the ISO data for these regions, but it could be worth while checking
whether they show the same spectral signature like WS95-80.

It may be significant in that context that all three regions which we
marked in Fig. \ref{iso_mbol} are star clusters which lie just on the
edge of a rather steep increase in CO flux.

Concerning the coincidence and offset between CO intensity and simulated
m$_{bol}$ (see caption of Fig. \ref{CO_mbol}), we see this in the context
of the age distribution in the Antennae: The overlap region, which hosts
the youngest clusters, shows a clear correlation between CO flux and
the location of the clusters (as traced by the bolometric luminosity),
which is not the case for the northwestern region. There, most of the
clusters are rather around 10 Myrs old, an age where they have blown
free of their natal dust clouds, which can explain the observed offset.

\subsection{Extinction evolution}

In Sec. \ref{extinction} we gave two pieces of evidence that clusters
are usually formed in high extinction regions, and have cleared away
this natal dust cocoon by the age of 8-9 Myrs. First evidence is the
correlation between high-extinction older clusters with high extinction
young star clusters. This, we interpret to mean that a cluster of around
10 Myrs or older which suffers high extinction is often not obscured
by its own dust cocoon any more, but just happens to be located in a
region of more recent star formation where the dust content is still
generally high. This effect is not seen for clusters below 8 Myrs, and
sets in around 9 Myrs, therefore the clusters on average seem to clear
away their natal cloud around that age.

The same result is obtained from Fig. \ref{AV_age_relation}, where -
despite large variations in A$_V$ within each age bin - the evolution of
extinction with age clearly declines.  Also here 8-9 Myrs is approximately
the limiting age above which many clusters have a sufficiently low
extinction that one can be sure that they have emerged from their natal
dust cocoon.

\subsection{Impact of extinction}

As shown in the previous section, owing to extinction, a careful analysis
of the optical data is required if sensible parameters, like for example
total masses and thereby star formation rates, are to be derived from
optical wavelengths alone in galaxies like the Antennae.

Here, a combination of considerably deeper optical images and a moderate
(A$_V \approx 1.3$) average extinction means that only a small fraction
of clusters which are detected in Ks is undetected in the other bands.
This holds even for U band (see Table \ref{photcompare}).

However, converting the observed cluster magnitudes into a quantity
like mass without information on the extinction leads to unacceptable
uncertainties. This is even the case if age information of the single
clusters is taken into account. In that case, the mass, assuming zero
extinction, will always be underestimated. In our case, only a quarter
of the mass are derived from U and B observations.

Even worse results are obtained if average ages and extinctions
are assumed, then for more or less reasonable assumptions for these
averages, the derived total mass can easily vary by orders of magnitude.
This behaviour will be less extreme in galaxies where the major star
formation event is not equally young, because after a few hundred
megayears, the luminosity evolution of a star cluster is much slower.

On the bright side, we showed in the previous section that it is possible
to obtain reasonable estimates of both, ages and extinction, of these
simple stellar populations by fitting the observed broadband colours with
these two free parameters. However, it is crucial there to have a broad
wavelength baseline, therefore optimally one needs a detection in both,
U and K band, or at least in one of them.

\subsection{Luminosity- and mass function}

As shown in Sec. \ref{photmass}, we constructed luminosity- and mass
functions for our Ks-detected clusters. Even though both functions can
in principle be fit by broken power laws with turnover cluster masses of
2.4 and 3 x 10$^5$M$_{\sun}$, respectively, we caution that both function
do not allow for this straightforward interpretation.

While the luminosity function is easily constructed and also completeness
correction is relatively easy, it is not very meaningful to compare this
luminosity function to, for example, that of globular cluster systems,
because the strong variation of mass-to-light ratios between the clusters
building up this luminosity function means that a given luminosity
cannot reliably be converted into a cluster mass. The observed ``turnover
mass'', where the two power laws which were fit to the function cross, is
converted into a characteristic mass assuming the mass-to-light ratio of
the median cluster mass in our sample, which is 9.2 Myrs, but particularly
the high luminosity clusters are often rather around 6 Myrs old, where
the mass-to-light ration differs from that at 9.2 Myrs substantially.

The mass function has a similar problem: While it is simple to construct a
mass function from those clusters which had ages and extinctions assigned
and therefore an individual mass estimate, it is difficult to apply a
completeness correction. Even though above $\approx$ 10$^5$M$_{\sun}$,
the mass function should be complete for all ages, the completeness limit
for the median age lies around 3x10$^4$M$_{\sun}$, but in the mass range
between these two values, the mass function will be affected by those
clusters which have ages where the completeness limit lies in that mass
range. And from the mass function strictly above 10$^5$M$_{\sun}$, one
cannot really reliably determine a turnover, expected to lie between 1
and a few times 10$^5$M$_{\sun}$.  Data which do not accurately probe the
turn-over provide very little constraint on the overall mass function.

So even though we constructed the luminosity- and mass functions, mainly
for the sake of comparison with what other authors found \citep[][for
the luminosity and mass function, respectively]{W99, ZhangFall}, and
find good agreement at least for the luminosity function, we caution
not to take these results too seriously. To improve the results, K-band
observations would be required which go $\approx$ 2.5 mag deeper, pushing
the completeness limit down to around 10$^4$M$_{\sun}$ for all ages.
And we note that without the near-infrared data, accurately accounting
for the extinction is challenging, making it difficult to estimate
robust total stellar masses for individual clusters.  

Even though this condition is fulfilled for the optical HST data alone
(from which the mass function in \cite{ZhangFall}, which could be fit by a
single power law, was constructed), we think that the broader wavelength
coverage obtained by including the K-band substantially improves the
age- and extinction-, and thereby the photometric mass determinations.

\section{Summary}\label{summary}

To summarize, our main conclusion from the analysis of VLT/NTT Antennae
near-infrared images, in combination with the HST optical images, is that
there is evidence that the majority of the bright, young cluster
population will not evolve into globular cluster, but rather disrupt on
timescales as short as tens of Myrs. If this was a generic property of
starbursts, the amount of star formation in similar events would have been
underestimated so far.

Large uncertainties are also introduced into estimates of star formation rates
if individual cluster ages and extinction are not taken into account.
In detail:

\begin{itemize}

\item Most Ks-band detected star clusters are young, the median age is
$\approx$ 9 Myrs, with a fairly narrow age range.  Most of the youngest
clusters (those below $\approx$ 5 Myrs) are found in the so-called
overlap region between the two galaxies. All other ages, also the few
older clusters (up to 200 Myrs) are relatively evenly distributed over
the two galaxies. The numbers of young clusters peak at two ages, 6 and
10 Myrs, the latter being mostly a selection effect (the clusters were
detected in K-band).

\item The most likely explanation for the narrow range of ages is
that clusters are destroyed or dissolve on scales of a few 10 Myrs.
An analysis of the cluster concentrations as a function of age does not
reveal a trend.  Thus it is unlikely that the clusters are born freely
expanding.  More likely is that they are marginally bound and stellar
evolutionary and dynamical mass loss lead to clusters becoming quickly
unbound.  Thus whether a cluster dissolves is likely to be dependent on
its mass distribution (concentration, etc) and initial mass function.

\item The extinction is variable, between A$_V\approx$0 and more than 10 mag
foreground screen extinction. The average value is 1.3 mag. It drops
with increasing cluster age, which is consistent with the picture that a
star cluster forms in a very dense and dust enshrouded environment and
blows free of this cocoon after a few Myrs. Most clusters have reached
this stage between the age of 8 and 9 Myrs.

\item Comparing photometry in the different bands, we note that
information about age and extinction for individual clusters is
necessary if meaningful conclusions about properties like total stellar
mass in clusters or (clustered) star formation rate are to be obtained.
Not taking into account this information, or assumption of average values
can lead to results differing by orders of magnitudes. However, both
age and extinction information can be obtained from broadband colours
alone if the wavelength coverage is broad enough.

\item A lower limit to the star formation rate over the last 25 Myrs is
$\approx$20 M$_{\sun}$ yr$^{-1}$, a relatively typical value compared to
those for other nearby starbursts.  But this is taking into account only
the detected clusters and not corrected for any potential dissolution of
clusters - including the latter, the lower limit for the SFR increases
to $\approx$50 M$_{\sun}$ yr$^{-1}$.

\item Even though the Ks-band luminosity- and mass functions are best fit
by broken power laws, and the turnover masses at face value lie around
(up to factor of 2) the characteristic mass of globular cluster mass
functions, this result should not be over-interpreted, due to limitations
implied by systematic uncertainties.

\end{itemize}

\begin{acknowledgements}
Part of SM's work was done at Leiden Observatory, The Netherlands,
financed by the EU RT Network ``Probing the Origin of the Extragalactic
Background Radiation'' and as a component of her Ph.D. thesis work at MPE.
We thank the ESO OPC for their generous allocation of observing time
and the staff of Paranal and La Silla for their support during the
observations presented here. Finally, we would like to thank the referee,
Peter Anders, for his insightful comments.

\end{acknowledgements}


\begin{thebibliography}{}

\bibitem[Anders and Fritze-v. Alvensleben (2003)]{AndersFvA03} Anders P., \&
Fritze-v. Alvensleben U., 2003, A\&A,  401, 1063


\bibitem[Anders et al. (2004)]{Anders04} Anders P., Bissantz N.,
Fritze-v. Alvensleben U., de Grijs R., 2004, MNRAS, 347, 196


\bibitem[Barnes \& Hernquist (1991)]{BH91} Barnes J.E. \& Hernquist L.E.,
1991, ApJ, 370, L65

\bibitem[Barnes (1988)]{Barnes88} Barnes J.E., 1988, ApJ, 331, 699

\bibitem[Barnes (2004)]{barnes04}Barnes, J. E. 2004, MNRAS, 350, 798

\bibitem[Bastian et al. (2005)]{Bastian05}Bastian N., Gieles M., Lamers H.J.G.L.M., 
Scheepmaker R.A., de Grijs R., 2005, A\&A, 431, 905

\bibitem[Boily \& Kroupa (2003)]{BoilyKroupa03} Boily C.M. \& Koupa P., 2003, MNRAS, 338, 665

\bibitem[de Marchi et al. (1997)]{deMarchi97} de Marchi G., et al, 1997, ApJL, 479, 27

\bibitem[Doyon et al. (1994)]{doyon94} Doyon, R., Joseph, R. D., \&  Wright,
G. S. 1994, ApJ, 421, 101

\bibitem[Fabbiano (2001)]{Fabbiano} Fabbiano G., Zezas A., Murray S.S,
2001, ApJ, 554, 1035

\bibitem[Fall \& Zhang (2001)]{FallZhang} Fall S.M. \& Zhang Q., 2001,
ApJ, 561, 751

\bibitem[Fall (2004)]{fall04}Fall, S. M. 2004, astro-ph/0405064

\bibitem[F\"orster-Schreiber et al. (2003)]{FoersteretalM82} F\"orster-Schreiber N.M., 
Genzel R., Lutz D., Sternberg A., 2003, ApJ, 599, 193

\bibitem[Kleinmann \& Hall (1986)]{KH86} Kleinmann S.G., Hall D.N.B.,
1986, ApJS, 62, 501

\bibitem[Kravtsov \& Gnedin (2003)] {KravtsovGnedin} Kravstov A.V., \&
Gnedin, O.Y., 2005, ApJ, 623, 650

\bibitem[Kroupa (2001)]{Kroupa2001} Kroupa P., 2001, MNRAS, 322, 231

\bibitem[Kunze et al. (1996)]{Ketal96}Kunze, D. et al. 1996, A\&A,
315, 101

\bibitem[Lada \& Lada (2003)]{LadaLada} Lada C.J. \& Lada E.A., 2003, ARA\&A, 41, 57

\bibitem[Lan\c{c}on \& Wood (2000)]{Lancon00} Lan\c{c}on, A. \& Wood,
P.R., 2000, A\&AS, 146, 217

\bibitem[Landolt (1983)]{Landolt83} Landolt, A. 1983, AJ, 88, 439

\bibitem[Leitherer et al. (1999)]{Letal99} Leitherer, C., et al.  1999,
ApJS, 123, 3

\bibitem[Mac Low \& Klessen (2004)]{maclow04}Mac Low, M.-M., \& Klessen,
R. S. 2004, Rev. Mod. Phys., 76, 126

\bibitem[Mathis, Rumpl \& Nordsieck (1977)]{Mathis77}Mathis J.S., Rumpl
W. \& Nordsieck K.H., 1977, ApJ, 217, 425

\bibitem[Mengel et al. (2001)]{Mengel01} Mengel, S., Lehnert, M.D.,
Thatte, N., Tacconi-Garman, L. E., \& Genzel,~R. 2001, ApJ, 550, 280

\bibitem[Mengel et al. (2002)]{Mengel02} Mengel S., Lehnert M.D., Thatte
N., \& Genzel R., 2002, A\&A, 383, 137

\bibitem[Meurer et al. (1995)]{Meurer95} Meurer, G.R., Heckman, T. M.,
Leitherer, C., Kinney, A., Robert, C., \& Garnett, D. R. 1995, AJ,
110, 2665

\bibitem[Mihos, Bothun, \& Richstone (1993)]{mihos93}Mihos, J. C.,
Bothun, G. D., \& Richstone, D. O. 1993, ApJ, 418, 82

\bibitem[Mirabel et al. (1998)]{Mirabel98} Mirabel L.F., Vigroux L.,
Charmandaris V., Sauvage M., Gallais P., Tran D., Cesarsky C., Madden
S.C., Duc P.-A., 1998, A\&A, 333, L1

\bibitem[Moffett \& Barnes(1979)]{MoffettBarnes79} Moffett T.J. \& Barnes T.G.III, 1979, AJ, 84, 627

\bibitem[Neff \& Ulvestad (2000)]{NeffUlvestad00} Neff S.G., \& Ulvestad J.S, 2000, AJ, 120, 670

\bibitem[Osterbrock (1974)]{Osterbrock74} Osterbrock D.E., 1974, {\it
Astrophysics of Gaseous Nebulae}, W.H. Freeman and Company

\bibitem[Origlia et al. (1998)]{Origlia98} Origlia L., Goldader J.D., Leitherer C., Schaerer D., \& Oliva E., 
1998, ApJ, 514, 96

\bibitem[Origlia \& Oliva (2000)]{OO00} Origlia L., Oliva E., 2000, A\&A, 357, 61

\bibitem[Schulz et al. (2002)]{Schulz02} Schulz J., Fritze-v. Alvensleben U., M\"oller C.S., \& Fricke K.J., 
2002, A\&A, 392, 1

\bibitem[Takahashi \& Portegies Zwart (2000)]{TPZ}Takahashi, K., \&
Portegies Zwart, S.  F.  2000, ApJ, 535, 759

\bibitem[Tremonti et al. (2001)]{tremonti01}Tremonti, C. A., Calzetti,
D., Leitherer, C., \& Heckman, T. M. 2001, ApJ, 555, 322

\bibitem[Vazquez \& Leitherer (2005)]{VazquezLeitherer05}Vazquez G.A. \& Leitherer C., 2005, ApJS, in press

\bibitem[Vesperini \& Zepf (2003)]{VesperiniZepf} Vesperini E., Zepf S.E., 2003, ApJL, 587, 97

\bibitem[Wang et al. (2004)]{Wangetal04} Wang Z., Fazio G.G., Ashby M.L.N., Huang J.-S., Pahre M.A.,
Smith H.A., Willner S.P., Forrest W.J, Pipher J.L., and Surace J.A., 2004, ApJS, 154, 193

\bibitem[Whitmore \& Schweizer (1995)]{WS95} Whitmore, B. \& Schweizer,
F. 1995, AJ, 109, 960

\bibitem[Whitmore et al. (1999)]{W99}Whitmore, B. C., Zhang, Q., Leitherer, C.
Fall, S. M., Schweizer, F., \& Miller, B. W. 1999, AJ, 118, 1551

\bibitem[Wilson et al. (2001)]{Wilson01} Wilson, C.  D., Scoville, N.,
Madden, S.  C., \& Charmandaris, V.  2000, ApJ, 542, 120

\bibitem[Zhang \& Fall (1999)]{ZhangFall} Zhang, Q. \&  Fall, S. M. 1999,
ApJ, 527, 81

\end{thebibliography}
\end{document}